\documentclass[12pt]{article}
\usepackage{epsf,amsfonts,amssymb,epsfig,amsmath,mathtools,graphics,slashed}
\usepackage{hyperref,hep,graphicx,subfig}
\usepackage{rotating}
\usepackage{xcolor}
\usepackage{verbatim}

\newcommand{\nn}{\notag \\}

\newcommand{\ves}{\varepsilon}

\makeatletter

\@addtoreset{equation}{section}
\makeatother

\begin{document}

\begin{titlepage}

\vfill


\vfill

\begin{center}
   \baselineskip=16pt
   {\Large\bf Critical Dynamics of Holographic Superfluids}
  \vskip 1.5cm
  \vskip 1.5cm
      Aristomenis Donos and Polydoros Kailidis\\
   \vskip .6cm
   \begin{small}
      \textit{Centre for Particle Theory and Department of Mathematical Sciences,\\ Durham University, Durham, DH1 3LE, U.K.}
   \end{small}\\            
\end{center}

\vfill

\begin{center}
\textbf{Abstract}
\end{center}
\begin{quote}
We study the nearly critical behaviour of holographic superfluids at finite temperature and chemical potential. Using analytic techniques in the bulk, we derive an effective theory for the long wavelength dynamics of gapless and pseudo-gapped modes, at first subleading order in a derivative expansion and we match the classical limit of our field theory construction in a companion paper. Specifically, we obtain the constitutive relations for the stress tensor and electric current, as well as a time evolution equation for the order parameter at next-to-leading order. In addition, we get explicit formulas for all the transport coefficients in terms of background quantities. We carry out numerical cross-checks with the predictions of our effective theory close to the critical point.
\end{quote}

\vfill

\end{titlepage}

\setcounter{equation}{0}

\section{Introduction}

The holographic duality \cite{Aharony:1999ti,Witten:1998qj} provides a powerful framework for studying strongly coupled systems at finite temperature and chemical potential \cite{Hartnoll:2016apf}. In a certain large $N$ limit, difficult questions about such systems are mapped to well-defined, exact computations in a classical theory of gravity of one dimension higher. Although the duality is, in principle, valid at all energy scales, one of its most fruitful applications is in analysing the hydrodynamic behaviour of thermal field theories, which is expected to be universal \cite{Policastro:2002se, Baier:2007ix, Bhattacharyya:2008jc}.

Broadly speaking, hydrodynamics \cite{Landau1987Fluid,Kovtun:2012rj, Romatschke:2017ejr} governs the collective behaviour of any finite temperature system at sufficiently late times and long wavelengths. In this regime, fluctuations of the high-energy degrees freedom are suppressed, and only the conserved charges are important. The time evolution is simply dictated by the corresponding conservation laws and the only non-trivial input of hydrodynamics is in expressing the conserved currents in terms of the conserved charges, in a derivative expansion scheme. Our ignorance about the microscopic theory is packaged in the coefficients of the hydrodynamic perturbative expansion, the transport coefficients.

Of particular interest is the low-energy effective dynamics of thermal states in the vicinity of a second order phase transition. The universal description of phase transitions, pioneered by Landau, involves the so-called order parameter which drives the transition, and becomes non-trivial below a critical temperature $T_c$. Close to the critical point, the thermal fluctuations of the order parameter give rise to non-mean field theory static phenomena leading to non-analyticities of thermodynamic quantities. Moreover, it crucially alters the late-time evolution of the theory and must be retained in any effective description, along with the conserved charges of the system \cite{RevModPhys.49.435, Folk_2006}. The reason is that the order parameter corresponds to an almost gapless degree of freedom, leading to the phenomenon of critical slowing down \cite{Stephanov:2017ghc}.

In this work, our focus will be on superfluid phase transitions. In the simplest possible setup, the role of the order parameter is played by a complex scalar operator, charged under a global $U(1)$ symmetry. Below $T_c$, this operator condenses and the continuous symmetry is spontaneously broken by the thermal state, resulting in the presence of a Goldstone mode in the low-energy spectrum. The Goldstone mode corresponds to fluctuations of the phase of the complex order parameter, whereas fluctuations of its amplitude correspond to the almost gapless mode near criticality, the Higgs/amplitude mode \cite{2015ARCMP...6..269P}. The complete hydrodynamic theory of superfluids deep in the broken phase, at first order in the derivative expansion,  was presented in \cite{Bhattacharya:2011tra}. On the other hand, close to $T_c$, the dynamics of the order parameter and the charge sector is captured by Model F in the classification of \cite{RevModPhys.49.435}. However, this model is accurate in a decoupling limit from the dynamics of the energy and momentum charges.

Superfluid states of matter were first realised holographically in \cite{Gubser:2008px} and \cite{Hartnoll:2008vx}. In the gravitational picture, the thermal state is modeled by a black hole with Hawking temperature $T$, corresponding to the field theory temperature. Below a certain $T=T_c$, a second branch of black hole solutions with scalar hair develops. Assuming that this new family of solutions has a lower free energy, a phase transition takes place. Since their introduction, holographic superfluids have been studied extensively. Most of the investigations of their hydrodynamic regime have been numerical in nature (see, e.g., \cite{Amado:2009ts, Arean:2021tks, Flory:2022uzp}), with the exception of some notable analytic treatments \cite{Herzog:2010vz}.

More recently, we proposed an analytic technique based on the Crnkovic-Witten symplectic current \cite{Crnkovic:1986ex}. This enables one to derive the linearised effective dynamics of the dual field theory, without having to solve the equations of motion in the bulk. An important byproduct of the method is that it provides us with analytic expressions for the transport coefficients in terms of background data. The technique has been applied to holographic superfluids both away from the critical point \cite{Donos:2022www}, as well as in its proximity \cite{Donos:2022xfd, Donos:2022qao}. In \cite{Donos:2022www}, the resulting theory agrees with the predictions of superfluid hydrodynamics in the absence of a background supercurrent \cite{Herzog:2011ec}. The investigation of \cite{Donos:2022qao}, based on the probe limit, where the dual stress tensor is decoupled from the fluctuations of the order parameter, produced a theory equivalent to Model F.

The goal of the present work is to apply the same technique to holographic superfluids near the critical point, including fluctuations of the metric in the bulk. From the dual mean field theory viewpoint, this will produce the complete effective dynamics near the phase transition, involving the order parameter and its coupling to all of the conserved charges of the system.

In a separate paper \cite{Donos:2025jxb}, we constructed the effective theory based on an independent approach, guided by hydrodynamic considerations \cite{Kovtun:2012rj}, as well as the Keldysh-Schwinger formalism for non-equilibrium systems \cite{Crossley:2015evo, Haehl:2015uoc, Liu:2018kfw}. Our effective theory can be viewed as a covariant generalisation of previous work by Khalatnikov and Lebedev \cite{KhalatnikovLebedev1978}, but it crucially involves an additional term in the order parameter equation, with complex coefficient $Z_\pi$, which was absent in \cite{KhalatnikovLebedev1978}. 

The findings of the current paper are in agreement with the constructions in \cite{Donos:2025jxb}. Specifically, we obtain the constitutive relations for the stress tensor and $U(1)$ current and a time evolution equation for the order parameter at next-to-leading order in a perturbative expansion scheme that fit nicely in the framework of \cite{Donos:2025jxb}. In addition, we obtain explicit formulas for all the transport coefficients, which obey the constraints proposed in \cite{Donos:2025jxb}, based on the entropy current and Onsager reciprocity. The coefficient $Z_\pi$ is found in terms of certain bulk integrals of background quantities, whereas the rest of the transport coefficients are expressed in terms of thermodynamics and horizon data.

This paper is organised as follows. In Section \ref{sec:setup} we introduce the class of holographic models we wish to study. In Section \ref{sec:Background_linear} we focus on backgrounds close to the phase transition critical point and discuss their thermodynamic fluctuations. In Section \ref{sec:Hydro_intro} we consider real-time fluctuations and analyse the perturbative structure of hydrodynamic fluctuations near criticality. We also summarise the main points of the symplectic current technique. In Section \ref{sec:Apply_SC}, we extract the relevant effective theory and present analytic expressions for the transport coefficients of the theory. In Section \ref{sec:Numerics} we check numerically the quasinormal mode spectrum of the effective theory by solving the exact equations of motion. Finally, in Section \ref{sec:Discuss}, we conclude with a discussion and outlook. Some technical details of the computation are gathered in the appendices for the interested reader.

\section{Holographic Setup}\label{sec:setup}

The basic ingredient of our bulk theory, in order to model a superfluid phase transition, is a complex scalar field $\psi$ that is dual to the boundary operator $\mathcal{O}_\psi$. The global $U(1)$ under which  $\mathcal{O}_\psi$ is charged, corresponds to a local symmetry in the bulk, gauged by the one-form $A_\mu$. The VEV of $\mathcal{O}_\psi$  will play the role of the order parameter in our system. Moreover, we will include a relevant operator $\mathcal{O}_\phi$ that introduces an additional deformation parameter $\phi_{s}$. 

The system is described by the bulk action\footnote{In this work, Greek letters are bulk indices corresponding to $t,r,x,y$,  Latin letters are boundary indices, with $a,b,c,...$  corresponding to $t,x,y$ and $i,j,k,...$ corresponding to the spatial directions $x,y$ only.},
\begin{align}\label{eq:bulk_action}
S_{bulk}=\int d^{4}x\,\sqrt{-g}\,\left(R-V(\phi,|\psi|^{2})-\frac{1}{2}\partial_{\mu}\phi\,\partial^{\mu}\phi-\frac{1}{2}(D_{\mu}\psi)(D^{\mu}\psi)^{\ast}-\frac{1}{4}\tau(\phi,|\psi|^{2})\,F^{\mu\nu}F_{\mu\nu} \right)\,.
\end{align}
The above action is invariant under the local gauge transformations $A\to A+d\Lambda$ and $\psi\to e^{-iq_e \Lambda}\,\psi$, given the covariant derivative $D_{\mu}\psi=\nabla_{\mu}\psi+iq_e A_{\mu}\,\psi$ and the field strength $F=dA$. In order for the bulk equations of motion to admit asymptotically $AdS_4$ solutions with unit radius, we will assume that the functions $V$  and $\tau$ admit the analytic expansion,
\begin{align}
V&\approx -6+\frac{1}{2}m_\psi^2\,|\psi|^2+\frac{1}{2}m_\phi^2\,\phi^2+\cdots\,,\nn
\tau&\approx 1+c_\psi\,|\psi|^2+c_\phi\,\phi+\cdots\,,
\end{align}
for small values of the scalar fields $\psi$ and $\phi$.

The superfluid phase of our system corresponds to a background with a non-trivial profile for the complex bulk scalar $\psi$. In this case, the field decomposition $\psi=\rho\,e^{iq_e\theta}$ is well defined, and we shall use this choice of variables, along with $B=A+d\theta$, from now on\footnote{The only exception to this is in Appendix \ref{app:Zpi}.}. We will also continue to refer to the one-form $B$ as the gauge field, in a slight abuse of terminology. In terms of the new field variables the bulk equations of motion derived from \eqref{eq:bulk_action} read,
\begin{align}\label{eq:eom}
R_{\mu\nu}-\frac{1}{2}g_{\mu\nu}V-\frac{\tau}{2}\left( F_{\mu\rho}F_{\nu}{}^{\rho}-\frac{1}{4}g_{\mu\nu}\,F^{2}\right) \qquad\qquad\qquad\qquad\qquad\qquad&\nn
-\frac{1}{2}\partial_{\mu}\phi\,\partial_{\nu}\phi-\frac{1}{2}\partial_{\mu}\rho\,\partial_{\nu}\rho-\frac{1}{2}q_e^2 \rho^{2}B_\mu\,B_\nu=&0\,,\nn
\nabla_{\mu}\nabla^{\mu}\phi-\partial_{\phi}V-\frac{1}{4}\partial_\phi\tau\,F^{2}=&0\,,\nn
\nabla_{\mu}\nabla^{\mu}\rho-2\,\rho\, \partial_{\rho^{2}}V-\frac{1}{2}\,\rho\,\partial_{\rho^{2}}\tau\,F^{2}-q_e^2\,\rho\,B^{2}=&0\,,\nn
\nabla_{\mu}(\tau\,F^{\mu\nu})-q_e^2\,\rho^{2}\,B^\nu=&0\,,
\nn
\nabla_{\mu}\left(\rho^{2}B^\mu\right)=&0\,.
\end{align}

In order to construct the thermal states of the normal as well as the broken phase of our system, we need to make the following ansatz,
\begin{align}\label{eq:background}
ds^{2}&=-U(r)\,dt^{2}+\frac{dr^{2}}{U(r)}+e^{2g(r)}\,\left(dx^{2}+dy^{2} \right)\,,\nn
B&=a(r)\,dt,\qquad \phi=\phi(r),\qquad \rho=\rho(r)\,,
\end{align}
which is sufficiently general to capture both states corresponding to the background bulk spacetime. The above choice of coordinates fixes the radial coordinate apart from a global shift which we will use to fix the event horizon position at $r=0$. Near the horizon, regularity implies the analytic expansions,
\begin{align}\label{eq:nh_bexp}
&U(r)\approx\,4\pi T\,r+\mathcal{O}(r^{2}),\qquad g(r)\approx g^{(0)}+g^{(1)}r+\mathcal{O}(r^2),\qquad a(r)\approx r\,a^{(0)}+\cdots \,,\nn
&\phi(r)\approx \phi^{(0)}+\mathcal{O}(r),\qquad \rho(r)\approx \rho^{(0)}+\mathcal{O}(r)\,,
\end{align}
where $T$ is the Hawking temperature.

On the other end of the geometry, close to the conformal boundary at $r\to\infty$, we impose the expansions,
\begin{align}\label{eq:nb_exp}
&U(r)\approx (r+R)^{2}+\cdots+\frac{g_{(v)}}{r+R}+\cdots,\qquad g(r)\approx \ln(r+R)+\mathcal{O}(r^{-1})\,,\nn
&a(r)\approx \mu-\frac{\varrho}{r+R}+\cdots\,,\nn
&\phi(r)\approx \phi_s\,(r+R)^{\Delta_{\phi}-3}+\cdots +\phi_v\,(r+R)^{-\Delta_{\phi}}+\cdots\,,\nn
&\rho(r)\approx \rho_s\,(r+R)^{\Delta_{\psi}-3}+\cdots+\rho_v\,(r+R)^{-\Delta_{\psi}}+\cdots\,,
\end{align}
where the conformal dimensions $\Delta_\psi$ and $\Delta_\phi$ of the dual operators $\mathcal{O}_\psi$ and $\mathcal{O}_\phi$ are fixed by the bulk masses according to $\Delta_\psi\,(\Delta_\psi-3)=m_\psi^2$ and $\Delta_\phi\,(\Delta_\phi-3)=m_\phi^2$.
Here $\rho_s$ and $\phi_s$ are the scalar operator sources  and $\mu$ is the chemical potential of the thermal state.  The global shift in the radial coordinate which fixes the horizon at $r=0$ is reflected by the constant of integration $R$. In this paper we will consider spontaneous breaking of the global $U(1)$ and for this reason we will fix $\rho_s=0$ for the background.  The field theory charge density is given by the constant of integration $\varrho$ and the expectation values of the field theory operators in the thermal state are fixed by\footnote{Without loss of generality, we will choose the phase of the complex bulk scalar to vanish in the background. Moreover, we normalise the expectation value of the complex scalar as $\frac{\delta S_{tot}}{\delta s_\psi}=\frac{\sqrt{-\gamma}}{2}\langle\mathcal{O}_\psi^\ast\rangle$, with $S_{tot}$ the bulk action plus the appropriate counterterms, $\gamma$ the determinant of the boundary metric and $s_\psi$ the source of the complex scalar.},
\begin{align}\label{eq:VEVs}
    \langle \mathcal{O}_\phi\rangle_b=(2\,\Delta_\phi-3)\phi_v\,,\quad \langle \mathcal{O}_\psi\rangle_b=(2\,\Delta_\psi-3)\rho_v\,.
\end{align}

The entropy density for this parametric family of backgrounds is given by $s=4\pi e^{2g^{(0)}}$, the horizon volume density. Another quantity that will prove useful later is the bulk charge density,
\begin{align}\label{eq:bulk_charge}
    \Phi(r)=\sqrt{-g}\,\tau\, F^{tr}\,.
\end{align}
Near the boundary, at $r=\infty$, it asymptotes to the field theory charge density $\varrho$. Close to the horizon, at $r=0$, it asymptotes to the horizon flux density $\varrho_h$,
\begin{align}
    \Phi(r\to \infty)\approx \varrho+\mathcal{O}\left((r+R)^{-2}\right),\quad \Phi(r\to 0)\approx \varrho_h+\mathcal {O}(r)=e^{2 g^{(0)}}\tau^{(0)}B^{(0)}_t+\mathcal {O}(r)\,.
\end{align}
Using the time component of the equation of motion of the gauge field in \eqref{eq:eom}, one can compute the radial derivative of the bulk charge density,
\begin{align}\label{eq:bulk_charge_1}
    \Phi'(r)=q_e^2\,\frac{e^{2g}}{U}\,a\,\rho^2\,.
\end{align}

The thermal states we will be interested in, are parametrised by the temperature $T$, the chemical potential $\mu$ and the scalar deformation parameter $\phi_s$. Moreover, we will consider a class of theories in which a second-order phase transition takes place. In such a model, for some fixed value of $\mu$ and $\phi_s$, there is a critical temperature $ T_c(\mu,\phi_s)$ below which we can find solutions with $\rho_s=0$ corresponding to the normal phase with $\langle \mathcal{O}_\psi \rangle_b=0$, as well as to the superfluid phase with non-trivial values for $\langle \mathcal{O}_\psi \rangle_b$. The hypersurface $\left(\mu,\phi_s,T_c(\mu,\phi_s)\right)$ defines a critical surface on which the free energy difference between the broken and normal phase is exactly zero.

For convenience, from now on we will choose our model such that it has conformal dimensions $\Delta_\psi=\Delta_\phi=2$ for our scalar operators. It would be useful to discuss the holographic renormalisation for our model \eqref{eq:bulk_action}. This would allow us to justify equations \eqref{eq:VEVs} and also extract the stress tensor and $U(1)$ current from the asymptotic expansions \eqref{eq:nb_exp}. For the more technical aspects of the procedure for this model we refer the reader to, e.g. \cite{Donos:2022www}, where we studied the superfluid phase of the system, away from the critical point.

For any solution to the equations of motion \eqref{eq:eom}, one can show that the gravitational and gauge field constraints in the bulk lead to the Ward identities,
\begin{align}\label{eq:contin}
   \nabla_a \langle T^{ab} \rangle=&F^{ba}\,\langle J_a \rangle+\frac{1}{2}\left(\langle \mathcal{O}_\psi \rangle^\ast \,D^b s_\psi +\langle \mathcal{O}_\psi \rangle\,D^b s_\psi^\ast\right)\,,\nn
   \nabla_a \langle J^a \rangle = &\frac{q_e}{2i}\left(\langle \mathcal{O}_\psi \rangle^\ast\,s_\psi-\langle \mathcal{O}_\psi \rangle\,s_\psi^\ast\right)\,.
\end{align}
for the boundary stress tensor and electric current, as well as the broken scale invariance equation
\begin{align}\label{eq:scale_inv}
\langle T^{ab}\rangle\gamma_{ab}=\phi_s\phi_v+\rho_s\rho_v\,.
\end{align}
Specifically for the background described by \eqref{eq:background}-\eqref{eq:nb_exp}, we can write the expectation values,
\begin{align}\label{eq:stress_bgr}
    \langle T^{tt}\rangle&=-2\,g_{(v)}-\phi_s\,\phi_v-\rho_s\,\rho_v=\epsilon\,,\nn\langle T^{ij}\rangle&=-\delta^{ij}\,g_{(v)}=\delta^{ij}\,p
\end{align}
where $\epsilon$ is the energy density and $p$ is the pressure of the thermal state. By evaluating the on-shell action and its first variation, we can show that these quantities obey,
\begin{align}\label{eq:thermo_rel}
    &\epsilon+p=T s+\mu \varrho\,,\nn
&dp=s\,dT+\varrho\,d\mu+\phi_{v}\,d\phi_s+\rho_v\, d\rho_s\,,
\end{align}
which are the standard thermodynamic relations between pressure and energy and the first law.

For later reference, it will also be useful to define various thermodynamic susceptibilities through variations of $\varrho,\,s,\,\phi_v,\,\rho_v$ as functions of $T,\,\mu,\,\rho_s,\,\phi_s$,
\begin{align}\label{eq:suscept_defs}
\delta\varrho&=\chi \,\delta\mu+\xi\,\delta T+\nu_{\mu\rho}\,\delta\rho_s+\nu_{\mu\phi}\,\delta\phi_s\,,\nn
\delta s&=\xi\,\delta\mu+\frac{c_\mu}{T}\,\delta T+\nu_{T\rho}\,\delta\rho_s+\nu_{T\phi}\,\delta\phi_s\,,\nn  \delta\rho_v&=\nu_{\mu\rho}\,\delta\mu+\nu_{T\rho}\,\delta T+\nu_{\rho\rho}\,\delta\rho_s+\nu_{\rho\phi}\,\delta\phi_s\,,\nn
\delta\phi_v&=\nu_{\mu\phi}\,\delta\mu+\nu_{T\phi}\,\delta T+\nu_{\rho\phi}\,\delta\rho_s+\nu_{\phi\phi}\,\delta\phi_s\,.
\end{align}
Notice that all susceptibilities are evaluated at $\rho_s=0$, in the broken phase. Finally, we will examine the implications of scaling symmetry, present in all theories with conformal symmetry. Combining \eqref{eq:stress_bgr} and \eqref{eq:thermo_rel} gives us the scaling anomaly equation,
\begin{align}\label{eq:trace}
    3p=Ts+\mu\varrho+\phi_s\,\phi_v+\rho_s\,\rho_v\,.
\end{align}
Upon varying \eqref{eq:trace} with respect to $\rho_s,\,\mu,\,T$ and setting $\rho_s=0$ at the end, we find respectively,
\begin{align}\label{eq:phi_susc}
    \nu_{\rho\phi}&=\frac{1}{\phi_s}(2\,\rho_v-T\, \nu_{T\rho}-\mu\,\nu_{\mu\rho})\,,\nn
    \nu_{\mu\phi}&=\frac{1}{\phi_s}(2\,\varrho-T\, \xi-\mu\,\chi)\,,\nn
    \nu_{T\phi}&=\frac{1}{\phi_s}(2\,s-c_\mu-\mu\,\xi)\,.
\end{align}

\section{Thermodynamic fluctuations near criticality}\label{sec:Background_linear}

Our main goal is to extract the effective theory close to the critical point. Following closely the treatment in \cite{Donos:2022qao,Donos:2022xfd}, out of the three dimensional family of backgrounds described in Section \ref{sec:setup}, we will need to focus on a curve $\left( T(\ves),\mu(\ves),\phi_s(\ves)\right)$ with parameter $\ves$, which originates from the critical point $(T_c(\mu,\phi_s),\mu,\phi_s)$  for $\ves=0$, and for $\ves>0$ it lies in the broken phase. For small values of $\ves$ we can expand the thermodynamic coordinates of this curve as,
\begin{align}\label{eq:eps_intro}
    T(\ves)&\approx T_c+\delta T_{\ast (2)}\frac{\ves^2}{2}+\mathcal{O}(\ves^4)\,,\nn
    \mu(\ves)&\approx \mu+\delta \mu_{\ast (2)}\frac{\ves^2}{2}+\mathcal{O}(\ves^4)\,,\nn
    \phi_s(\ves)&\approx \phi_s+\delta \phi_{s\ast (2)}\frac{\ves^2}{2}+\mathcal{O}(\ves^4)\,.
\end{align}

The superfluid phase background along this curve can also be expanded in $\ves$ according to,
\begin{align}\label{eq:broken_exp}
U&=U_c+\frac{\varepsilon^2}{2}\,\delta U_{\ast(2)}+\frac{\varepsilon^4}{4!}\,\delta U_{\ast(4)}+\cdots,\quad g=g_c+\frac{\varepsilon^2}{2}\,\delta g_{\ast(2)}+\frac{\varepsilon^4}{4!}\,\delta g_{\ast(4)}+\cdots\,,\nn
 a&=a_c+\frac{\varepsilon^2}{2}\,\delta a_{\ast(2)}+\frac{\varepsilon^4}{4!}\delta a_{\ast(4)}+\cdots\,, \quad \phi=\phi_c+\frac{\varepsilon^2}{2}\,\delta \phi_{\ast(2)}+\frac{\varepsilon^4}{4!}\,\delta \phi_{\ast(4)}+\cdots\,,\nn
\rho&=\varepsilon\,\delta\rho_{\ast(0)}+\frac{\varepsilon^3}{3!}\,\delta\rho_{\ast(2)}+\cdots\,.
\end{align}
As expected, exactly at the critical point at $\ves=0$, the amplitude field vanishes. Right at the critical point, one can find a source-free static solution of the linearised amplitude equation. This solution, denoted by $\delta\rho_{\ast(0)}$, is the critical mode which drives the second-order phase transition from the bulk point of view. Following this mode takes us from the normal fluid phase to the thermodynamically preferred superfluid phase. Its non-linear effects through backreaction become important as we leave the critical point and go deeper in the broken phase. This justifies that all other fields depend quadratically on $\ves$.

Given the expansion above for the bulk fields near the critical point, we can read off the expansion of entropy density $s$ and the charge densities we defined earlier,
\begin{align}
    s&=s_c+\delta s_{\ast (2)}\frac{\ves^2}{2}+\mathcal{O}(\ves^4)\,,\quad\varrho=\varrho_c+\delta\varrho_{\ast (2)}\frac{\ves^2}{2}+\mathcal{O}(\ves^4)\,,\nn
    \varrho_h&=\varrho_c+\delta\varrho_{\ast h (2)}\frac{\ves^2}{2}+\mathcal{O}(\ves^4)\,,\quad
    \Phi(r)=\varrho_c+\delta \Phi_{b(2)}\, \frac{\ves^2}{2}+\mathcal{O}(\ves^4)\,,
\end{align}
where the bulk flux expansion coefficient $\delta \Phi_{b(2)}$ obeys the radial equation,
\begin{align}\label{eq:bulk_charge_der}
 \delta \Phi_{b(2)}'(r)=2\,q_e^2\frac{e^{2g_c}}{U_c}\,a_c\,\delta\rho_{\ast (0)}^2\,,
\end{align}
as can be seen by expanding equation \eqref{eq:bulk_charge_1} in $\ves$. For later convenience, it will be useful to introduce the difference between the boundary and horizon gauge field fluxes,
\begin{align}
    \Delta\varrho_h=\varrho-\varrho_h=\Delta\varrho_{h(2)}\frac{\ves^2}{2}+\mathcal{O}(\ves^4)\,,
\end{align}
with $\Delta\varrho_{h(2)}=\delta\varrho_{\ast(2)}-\delta\varrho_{\ast h(2)}$.

The effective theory we wish to construct captures the dynamics of certain long wavelength perturbations on top of the near critical background \eqref{eq:broken_exp}, from the boundary point of view. Moreover, by varying the thermodynamic parameters of our backgrounds we can study static linear fluctuations and these will be also useful later for the symplectic current method. All but one of these perturbations that we will need are described in \cite{Donos:2022www}. Here we will only enumerate them, referring the reader to \cite{Donos:2022www} for further details. It is important to note that, similar to the background solution, the static perturbations have to be expanded in powers of $\ves$ as well.

Following the discussion in \cite{Donos:2022www} (see also Subsection \ref{sec:space_time_perts} in this paper) it is convenient to introduce a function $S(r)$ behaving as $S(r) \approx \frac{\mathrm{ln} r}{4\pi T}+\cdots$, close to the horizon. In this way, the combination $t+S(r)$ becomes regular on the black hole horizon. We will also choose the asymptotics of $S(r)$ to approach $\mathcal{O}(1/r^4)$ close to the boundary, convenient for the holographic dictionary. It is also worth noting that $S$  follows our expansion scheme, $S=S_{[0]}+\mathcal{O}(\ves^2)$, since it depends on the temperature.

The first two thermodynamic perturbations emerge by varying the temperature and chemical potential of the background, respectively. More concretely, temperature variations yield the bulk fluctuations,\footnote{All bulk fields not appearing are understood to be zero. Note that $\delta_T B_r$ was omitted in \cite{Donos:2022www} due to a typo. }
\begin{align}\label{eq:static_pert_dT}
\delta_T g_{tt}&=-\partial_T U_{[0]}+\mathcal{O}(\ves^2)\,,\quad \delta_T g_{rr}=-\frac{\partial_T U_{[0]}}{U_c^2}+\mathcal{O}(\ves^2)\,,\quad \delta_T g_{tr}=U_c\,\partial_T S^\prime_{[0]}+\mathcal{O}(\ves^2)\,,\nn
\delta_T g_{ij}&=2\,\delta_{ij}\,e^{2g_c}\,\partial_T g_{[0]}+\mathcal{O}(\ves^2)\,,\quad \delta_T B_t=\partial_T a_{[0]}+\mathcal{O}(\ves^2)\,,\quad\delta_T B_r=-a_c\, \partial_T S'_{[0]}+\mathcal{O}(\ves^2)\,,\nn
\delta_T\phi&=\partial_T\phi_{[0]}+\mathcal{O}(\ves^2)\,,\quad \delta_T\rho=\frac{1}{\ves}\left(\partial_T\rho_{[0]}+\partial_T\rho_{[2]}\,\ves^2+\mathcal{O}(\ves^4)\right)\,,
\end{align}
while chemical potential variations give,
\begin{align}\label{eq:static_pert_dmut}
\delta_{\mu} g_{tt}&=-\partial_{\mu} U_{[0]}+\mathcal{O}(\ves^2)\,,\quad \delta_{\mu} g_{rr}=-\frac{\partial_{\mu} U_{[0]}}{U_c^2}+\mathcal{O}(\ves^2)\,,\quad \delta_{\mu} g_{ij}=2\,\delta_{ij}\,e^{2g_c}\,\partial_{\mu} g_{[0]}+\mathcal{O}(\ves^2)\,,\nn
\delta_{\mu} B_t&=\partial_{\mu} a_{[0]}+\mathcal{O}(\ves^2)\,,\quad \delta_{\mu}\phi=\partial_{\mu}\phi_{[0]}+\mathcal{O}(\ves^2)\,,\quad \delta_{\mu}\rho=\frac{1}{\ves}\left(\partial_{\mu}\rho_{[0]}+\partial_{\mu}\rho_{[2]}\,\ves^2+\mathcal{O}(\ves^4)\right)\,.
\end{align}
The overall scale in the above two perturbations is fixed so that $\delta_T g_{tt}=-4\pi\, r+\mathcal{O}(r^2)$, near the horizon, and $\delta_\mu B_t=1+\mathcal{O}(\frac{1}{r+R})$, near the boundary.

At this point, let us briefly comment on the notation. In the above, we have expanded the derivatives of the background fields with respect to $T$ and $\mu$ along the curve \eqref{eq:eps_intro}. For instance, $\partial_TU=\partial_TU_{[0]}+\partial_TU_{[2]}\,\ves^2+\cdots,$ and $\partial_T\rho=\frac{1}{\ves}\left(\partial_T\rho_{[0]}+\partial_T\rho_{[2]}\,\ves^2+\cdots\right)$. Notice that the factors $\partial_TU_{[2n]}, \partial_T \rho_{[2n]}, etc$ are order $\ves^0$. Comparing with the expansions \eqref{eq:broken_exp} we can identify e.g. $\delta U_{\ast(2)}=\partial_TU_{[0]}\delta T_{\ast(2)}+\partial_\mu U_{[0]}\delta \mu_{\ast(2)}+\partial_{\phi_s} U_{[0]}\delta \phi_{s\ast(2)}$. In addition, in the following, if $\delta_{st.}\in \{\delta_T,\,\delta_\mu,\,\delta_{\rho_s},\,\delta_{v_i},\,\delta_{m_i},\,\delta_{s_{ab}}\}$ and $f$ is one of the radial functions parameterising the bulk fields, then $\delta_{st.} f_{[2n]}$ will denote the $n+1$-$th$ term in the $\ves$ expansion of $\delta_{st.} f$, including factors of $\ves$. 

In thermal equilibrium, the fluid velocity $v^i$ defines the rest frame of the normal fluid. Its variation in the rest frame of the normal fluid gives rise to the gravitational perturbation in the bulk,
\begin{align} \label{eq:static_pert_dv}
\delta_{v_j} g_{ti}&=\delta^j_i\,\left(U_c-e^{2g_c}+\left(\frac{\delta U_{\ast(2)}}{2}-e^{2g_c}\,\delta g_{\ast(2)}+\mu\,\delta {f_g}_{[2]})\right)\ves^2+\mathcal{O}(\ves^4)\right)\,,\nn \delta_{v_j} g_{ri}&=\delta^j_i\,\left(-e^{2g_c}\,S_{[0]}^\prime +\mu\,\delta {f_c}_{[0]}+\mathcal{O}(\ves^2)\right)\,,\nn
\delta_{v_j}B_i&=\delta^j_i\,\left(\mu-a_c+\,\left(\frac{\delta\mu_{\ast(2)}}{2}+\mu\,\delta {f_b}_{[2]}-\frac{\delta a_{\ast(2)}}{2}\right)\ves^2+\mathcal{O}(\ves^4) \right)\,.
\end{align}

The most general superfluid thermal state can include a persistent super-current. Even though we are not including it in our backgrounds, it is important to consider the variation of our thermal states with respect to its gauge invariant source $m_j$. This is very similar to the fact that we have introduced the source $\rho_s$ and its variations but in the end we will set it equal to zero for our thermal states. The supercurrent source variations lead to the bulk fluctuations\footnote{The $\delta_{m_j}g_{ri}$ was also omitted in \cite{Donos:2022www} by mistake.},
\begin{align} \label{eq:static_pert_dmui}
\delta_{m_j} g_{ti}&=\delta^j_i\left( \,\delta {f_g}_{[2]}\,\ves^2+\mathcal{O}(\ves^4)\right)\,,\quad \delta_{m_j}B_i=\delta^j_i\left(1+\delta {f_b}_{[2]}\,\ves^2+\mathcal{O}(\ves^4)\right)\,, \nn \delta_{m_j}g_{ri}&=\delta^j_i\,\left( \delta {f_c}_{[0]}+\mathcal{O}(\ves^2)\right)\,.
\end{align}
The three bulk functions introduced above, near the horizon behave as,
\begin{align}
\delta {f_g}_{[2]}(r)=\delta {f_g}_{[2]}^{(1)}\,r+\cdots\,,\quad \delta {f_b}_{[2]}(r)=\delta {f_b}_{[2]}^{(0)}+\cdots\,, \quad  \delta {f_c}_{[0]}(r)=\delta {f_c}_{[0]}^{(0)}+\cdots\,,
\end{align}
and near the conformal boundary,
\begin{align}
\delta {f_g}_{[2]}(r)=\frac{1}{3}\,\frac{\mu\,\chi_{JJ}^{[2]}}{r+R}+\cdots\,,\quad \delta {f_b}_{[2]}=-\frac{\chi_{JJ}^{[2]}}{r+R}+\cdots\,, \quad \delta {f_c}_{[0]}=\mathcal{O}\left((r+R)^{-3}\right)\,,
\end{align}
with,
\begin{align}\label{eq:chiJJ}
  \chi_{JJ}=\chi_{JJ}^{[2]}\,\ves^2+\mathcal{O}(\ves^4)  
\end{align}
being the current-current susceptibility. Using the gauge field equation, it is straightforward to show that,
\begin{align}\label{eq:chi_JJ_rel}
    \chi_{JJ}^{[2]}=q_e^2 \int^\infty_0 dr\, \delta \rho_{\ast(0)}^2\,.
\end{align}

In addition to the thermodynamic perturbations we discussed above, we can generate a set of spacetime-independent perturbations by varying the static sources for the external metric. As explained in\cite{Donos:2022www}, they can be constructed by large diffeomorphisms in the bulk, which, from the field theory point of view, induce a change in the field theory metric:
$\delta \gamma_{ab}=2\,\delta s_{\left(ab\right)}$\,. The bulk perturbation corresponding to the source $\delta s_{tt}$ is given by,
\begin{align}
\delta_{s_{tt}}g_{tt}&=2\,U_c-\mu\,\partial_{\mu}U_{[0]}+\mathcal{O}(\ves^2)\,,\quad \delta_{s_{tt}}g_{tr}=U_c\,S^\prime_{[0]}+\mathcal{O}(\ves^2)\,,\nn
\delta_{s_{tt}}g_{rr}&=-\mu\,\frac{\partial_{\mu}U_{[0]}}{U_c^2}+\mathcal{O}(\ves^2)\,,\quad \delta_{s_{tt}}g_{ij}=2\,\delta_{ij}\,\mu\,e^{2g_c}\,\partial_{\mu}g_{[0]}+\mathcal{O}(\ves^2)\,,\nn
\delta_{s_{tt}}B_{t}&=-a_c+\mu\,\partial_{\mu}a_{[0]}+\mathcal{O}(\ves^2)\,,\quad \delta_{s_{tt}}B_{r}=-a_c\,S^\prime_{[0]}+\mathcal{O}(\ves^2)\,,\nn
\delta_{ s_{tt}}\phi&=\mu\,\partial_{\mu}\phi_{[0]}+\mathcal{O}(\ves^2)\,,\quad \delta_{s_{tt}}\rho=\frac{1}{\ves}\left(\mu\,\partial_{\mu}\rho_{[0]}+\mathcal{O}(\ves^2)\right)\,.
\end{align}
For the rest of the perturbations of this kind, we can write,
\begin{align}\label{eq:static_pert_source}
\delta_{s_{tj}}g_{ti}&=\delta^j_i\,\left(U_c+\left(\frac{\delta U_{\ast(2)}}{2}+\mu\,\delta {f_g}_{[2]}\right)\ves^2+\mathcal{O}(\ves^4)\right)\,,\nn \delta_{s_{tj}}B_i&=\delta^j_i\,\left(-a_c+\mu+\left(-\frac{\delta a_{\ast(2)}}{2}+\frac{\delta\mu_{\ast(2)}}{2}+\mu\,\delta {f_b}_{[2]}\right)\ves^2+\mathcal{O}(\ves^4)\right)\,,\nn
\delta_{s_{tj}}g_{ri}&=\delta^j_i\left(\mu\, \delta {f_c}_{[0]}+\left(\frac{\delta\mu_{\ast(2)}}{2} \delta {f_c}_{[0]}+\mu\, \delta {f_c}_{[2]}\right)\ves^2+\mathcal{O}(\ves^4)\right) \,,\nn
\delta_{s_{jt}}g_{ti}&=\delta^j_i\,e^{2g_c}+\mathcal{O}(\ves^2)\,,\quad \delta_{s_{jt}}g_{ri}=\delta^j_i\,e^{2g_c}\,S^\prime_{[0]}+\mathcal{O}(\ves^2)\,,\nn
\delta_{s_{ij}}g_{kl}&=2\,\delta^{\left(i\right.}_k \delta^{\left.j\right)}_l\,e^{2g_c}+\mathcal{O}(\ves^2)\,.
\end{align}

For the case of superfluids away from the critical point \cite{Donos:2022www}, the set of fluctuations generated by the variations $\{\delta_T,\,\delta_\mu,\,\delta_{m_i},\,\delta_{v_i},\,\delta_{s_{ab}}\}$ was sufficient to construct the leading part of the complete set of hydrodynamic fluctuations. Near criticality, the amplitude of the order parameter offers an additional long lived mode, as we discussed earlier. In order to generate the leading part of the relevant hydrodynamic excitation, the extra ingredient we need to introduce, is a thermodynamic perturbation carrying a source $\delta\rho_s$ for the amplitude mode\footnote{In \cite{Donos:2022xfd} and \cite{Donos:2022qao}, instead of the set $\{\delta_T\,,\delta_\mu\,,\delta_{\rho_s}\}$ of static linear fluctuations, a different one was used, namely $\{\delta_\ast\,,\delta_\#\}$, at leading order in the $\ves$ expansion. These linear fluctuations naturally form a three-dimensional vector space and the choice of perturbations is merely a choice of basis.},
\begin{align}\label{eq:static_pert_drho}
\delta_{\rho_s} g_{tt}&=\frac{1}{\ves}\left(-\partial_{\rho_s} U_{[0]}+\mathcal{O}(\ves^2)\right),\, \delta_{\rho_s} g_{rr}=\frac{1}{\ves}\left(-\frac{\partial_{\rho_s} U_{[0]}}{U_c^2}+\mathcal{O}(\ves^2)\right),\,
\delta_{\rho_s} g_{ij}=\frac{1}{\ves}\left(2\,\delta_{ij}\,e^{2g_c}\,\partial_{\rho_s} g_{[0]}+\mathcal{O}(\ves^2)\right)\,,\nn \delta_{\rho_s} B_t&=\frac{1}{\ves}\left(\partial_{\rho_s} a_{[0]}+\mathcal{O}(\ves^2)\right)\,,\quad
\delta_{\rho_s}\phi=\frac{1}{\ves}\left(\partial_{\rho_s}\phi_{[0]}+\mathcal{O}(\ves^2)\right)\,,\quad \delta_{\rho_s}\rho=\frac{1}{\ves^2}\left(\partial_{\rho_s}\rho_{[0]}+\partial_{\rho_s}\rho_{[2]}\,\ves^2+\mathcal{O}(\ves^4)\right)\,.
\end{align}
Notice that the perturbation generated by $\delta_{\rho_s}$  is sourceless at leading order and carries a source at order $\ves^0$, equal to 1, as encoded in the asymptotic behaviour,
\begin{align}\label{eq:source_static}
    \partial_{\rho_s}\rho_{[2]}\approx (r+R)^{\Delta_\psi-3}+\cdots+\partial_{\rho_s}\rho_{[2]}^v(r+R)^{-\Delta_\psi}+\cdots\,.
\end{align}

It is useful to expand the susceptibilities defined in \eqref{eq:suscept_defs}, the horizon charge densities and the variations of the scalar fields values at the horizon in $\ves$. In general we can write an expansion of the form,
\begin{align}\label{eq:susc_scale}
    c=\frac{1}{\ves^{n_c}}\left(c_{[0]}+c_{[2]}\ves^2+\mathcal{O}(\ves^4)\right)\,,
\end{align}
with $c$ any one of these quantities. For $\chi$, $\xi$, $c_\mu$, $\delta_T\varrho_h\equiv \partial_T\varrho_h$, $\delta_\mu\varrho_h\equiv \partial_\mu\varrho_h$ and $\nu_{\mu\phi}$, $\nu_{T\phi}$, $\nu_{\phi\phi}$, $\partial_T\phi^{(0)}$, $\partial_\mu\phi^{(0)}$ we have $n_c=0$. For $\nu_{T\rho}$, $\nu_{\mu\rho}$, $\delta_{\rho_s}\varrho_h\equiv\partial_{\rho_s}\varrho_h$, $\nu_{\rho\phi}$, $\partial_T\rho^{(0)}$,  $\partial_\mu\rho^{(0)}$, $\partial_{\rho_s}\phi^{(0)}$, we have $n_c=1$ and for $\nu_{\rho\rho}$, $\partial_{\rho_s}\rho^{(0)}$, $n_c=2$.

By examining the equations of motion for the perturbations generated by the variations $\{\delta_T\,,\delta_\mu\,,\delta_{\rho_s}\}$, we can deduce that the amplitude perturbation at leading order, in each of them, is a multiple of the critical mode everywhere in the bulk,
\begin{align}\label{eq:mult_rho}
    \partial_T\rho_{[0]}(r)&=\frac{\nu_{T\rho}^{[0]}}{\delta\rho_{\ast(0)}^v}\delta\rho_{\ast(0)}(r)\,,\quad\partial_\mu\rho_{[0]}(r)=\frac{\nu_{\mu\rho}^{[0]}}{\delta\rho_{\ast(0)}^v}\delta\rho_{\ast(0)}(r)\,, \quad \partial_{\rho_s}\rho_{[0]}(r)=\frac{\nu_{\rho\rho}^{[0]}}{\delta\rho_{\ast(0)}^v}\delta\rho_{\ast(0)}(r)\,.
\end{align}
As shown in Appendix \ref{app:susc_rel}, this observation yields several relations among the leading parts of the quantities appearing in \eqref{eq:susc_scale}.

\section{Hydrodynamic perturbations and the symplectic current}\label{sec:Hydro_intro}

In this section, we will outline the necessary ingredients to construct the hydrodynamic perturbations from a bulk point of view. In Subsection \ref{sec:space_time_perts} we discuss general aspects of spacetime-dependent perturbations and then, in Subsection \ref{sec:build_hydro_mode}, we move the focus to the perturbative structure of a generic hydrodynamic fluctuation. Finally, in Subsection \ref{sec:sympl_curr} we sketch the main steps of the symplectic current technique on which our construction relies.

\subsection{Spacetime-dependent perturbations}\label{sec:space_time_perts}

For the class of perturbations we wish to study, for any bulk field $\delta\mathcal{F}$ we can perform the Fourier mode decomposition,  
\begin{align}\label{eq:fourier_modes}
\delta\mathcal{F}(t,x^i;r)=e^{-i\, w\,(t+S(r))+iq_i x^i}\,\delta f(r)\,,
\end{align}
due to the translational symmetry of the background spacetime \eqref{eq:background}. This reduces the problem to a system of linear ODEs for the radial functions $\delta f(r)$. The function $S(r)$ is the same as the one we introduced in the previous section. Its behaviour near the horizon guarantees the correct infalling boundary conditions (provided that $\delta f(r)$ remains analytic there), whereas its behaviour close to the conformal boundary ensures that the boundary theory information is packaged entirely in the asymptotics of $\delta f(r)$.

Close to the boundary, the radial functions that parametrise our spacetime dependent perturbations will behave according to,
\begin{align}\label{eq:pert_uv_bcs}
\delta g_{ab}(r)&=(r+R)^2\,\left(\delta \gamma_{ab}+\cdots +\frac{\delta t_{ab}}{(r+R)^3}+\cdots\right)\,,\nn
\delta g_{ra}(r)&=\mathcal{O}\left(\left(r+R\right)^{-3}\right),\quad\delta g_{rr}(r)=\frac{\delta g_{rr}^v}{\left(r+R\right)^5}+\cdots\,,\nn
\delta B_a(r)&=\partial_a \delta\theta_{(s)}(r+R)+\delta s_a+\partial_a\delta\theta_{(v)}+\frac{\delta j_a}{r+R}+\cdots\,,\nn
\delta B_r(r)&=\delta\theta_{(s)}+\frac{\delta B_r^s}{(r+R)^2}+\cdots\,,\nn
\delta\phi(r)&=\frac{\delta\phi_{(v)}}{(r+R)^2}+\cdots,\quad \delta\rho(r)=\frac{\delta\rho_{(s)}}{r+R}+\frac{\delta\rho_{(v)}}{(r+R)^2}+\cdots\,.
\end{align}
The asymptotics of the functions $\delta g_{r\mu}$ can be fixed by using our freedom to choose coordinates. We stress that we do not choose to work in a particular coordinate system in the bulk, we only fix our coordinates asymptotically through the decays of the metric components $\delta g_{r\mu}$. Our choice for the fall-off of $\delta g_{r\mu}$ is such that it is consistent with the fall-off of  $\delta g_{r\mu}$ of the static perturbations described in the previous section, and this is the reason for keeping the pure-gauge term $\delta g_{rr}^v$ in \eqref{eq:pert_uv_bcs}\footnote{Adding to our perturbation a pure-gauge perturbation generated by the bulk diffeomorphism $r\to r+\xi^r$, with $\xi^r=\frac{\delta g_{rr}^v}{6}\frac{1}{(r+R)^2}+\cdots$ near the boundary, cancels the $\mathcal{O}(1/r^5)$ term from the asymptotic expansion of $\delta g_{rr}$ and shifts $\delta t_{ab}\to \delta t_{ab}+\eta_{ab}\frac{\delta g_{rr}^v}{3}$.}. The rest of the above asymptotic behaviour is fixed by the equations of motion of the bulk theory. In particular, the constant $\delta B_r^s$ is fixed in terms of the sources $\delta\rho_{(s)}$, $\delta\theta_{(s)}$ through the gauge field equation of motion \eqref{eq:eom}. It will later become convenient to introduce the gauge invariant combination, $\delta m_a=\delta s_a+\partial_a \delta\theta_{(v)}$ between the external perturbative source for the gauge field and the phase of the order parameter. 

The expectation values of the field theory stress tensor and $U(1)$ current are determined in terms of the integration constants of the near boundary expansions \eqref{eq:pert_uv_bcs} according to\footnote{We raise indices according to the relations $\delta t^{ab}=\eta^{ac}\eta^{bd}\delta t_{cd},\,\delta \gamma^{ab}=\eta^{ac}\eta^{bd}\delta \gamma_{cd},\,\delta j^a=\eta^{ab}\delta j_b$.}, 
\begin{align}\label{eq:stress_curr_uv}
    \delta\langle T^{ab}\rangle&=3\,\delta t^{ab}-3\,\eta^{ab}\eta^{cd}\,\delta t_{cd}+\left(g_{(v)}+\phi_s\phi_v\right)\delta \gamma^{ab}-\left(3\,g_{(v)}+\phi_s\phi_v\right)2\,\delta^{(a}_t\delta \gamma^{b)t}\nn&-\eta^{ab}\left(3\,g_{(v)}\delta \gamma_{tt}+\phi_s\phi_v\delta^{ij}\delta \gamma_{ij}+\delta\rho_{(s)}\rho_v+\phi_s \delta\phi_{(v)}+2\,\delta g_{rr}^v\right)\,,\nn
    \delta \langle J^a\rangle&=\delta j^a+\varrho\, \delta \gamma^{at}+\eta^{ab}\partial_b\,\delta B_r^s\,.
\end{align}

The expectation value and the external source for the complex scalar are parametrised according to,
\begin{align}
    \delta \langle \mathcal{O}_\psi\rangle&=\left(2\,\Delta_\psi-3\right)\left(\delta\rho_{(v)}+i\,q_e\,\rho_{v}\,\delta\theta_{(v)}\right)e^{-i\, w\,t+iq_i x^i}\,,\nn
    \delta s_\psi&= \left(\delta\rho_{(s)}+i\,q_e\,\rho_{v}\delta\theta_{(s)}\right)e^{-i\, w\,t+iq_i x^i}\,.
\end{align}

From the field theory point of view, the expectation values  $\delta\langle T^{ab}\rangle,\, \delta \langle J^a\rangle $ must obey the linearised Ward identities \eqref{eq:contin} and the trace condition \eqref{eq:scale_inv}. From the gravitational point of view, the constants $\delta t_{ab}$ and $\delta j_a$ are not independent, but are constrained by the gravitational and gauge field constraints when expanded in the radial coordinate close to the boundary. More specifically, the $ra$ and $rr$ components of Einstein's equations yield the stress tensor continuity\footnote{The linearised Christoffel symbols read: $\delta \Gamma^a_{bc}=\frac{\eta^{ad}}{2}\left(\partial_b \delta \gamma_{dc}+\partial_c\delta \gamma_{db}-\partial_d\delta \gamma_{bc}\right)$.} and the trace anomaly,
\begin{align}
    &\partial_a \delta\langle T^{ab}\rangle+\delta\Gamma^c_{ca}\langle T^{ab}\rangle+ \delta\Gamma^b_{ca}\langle T^{ca}\rangle=\eta^{bi}\varrho\left(\partial_i \delta s_t-\partial_t\delta s_i\right)-\eta^{bt}\mu\, q_e^2\rho_v^2\, \delta\theta_{(s)}+\eta^{bc}\partial_c\delta\rho_{(s)}\rho_v\,,\notag\\
    &\delta\langle T^{ab}\rangle \eta_{ab}+\langle T^{ab}\rangle\delta \gamma_{ab}=\phi_s\delta\phi_{(v)}+\delta\rho_{(s)}\rho_v\,.
\end{align}
Finally, the $r$ component of the gauge field equation close to the boundary leads to the current continuity,
\begin{align}
\partial_a\delta\langle J^a\rangle+\delta\Gamma^a_{at}\,\varrho=q_e^2\rho_v^2\, \delta\theta_{(s)}\,.
\end{align}

On the other end of our domain, close to the black hole horizon at $r=0$, we need to impose ingoing boundary conditions which we can achieve through the asymptotics,
\begin{align}\label{eq:gen_exp}
\delta g_{tt}(r)&= 4\pi T\,r\, \delta g_{tt}^{(0)}+\cdots\,,\quad
\delta g_{rr}(r)=\frac{\delta g_{rr}^{(0)}}{4\pi T\,r}+\cdots\, \,,\nn
\delta g_{ti}(r)&=\delta g_{ti}^{(0)}+r\,\delta g_{ti}^{(1)}+\cdots\,,\quad 
\delta g_{ri}(r)=\frac{\delta g_{ri}^{(0)}}{4\pi T\,r}+\delta g_{ri}^{(1)}+\cdots\,,\nn
\delta g_{ij}(r)&=\delta g_{ij}^{(0)}+\cdots\,,\quad\quad
\delta g_{tr}(r)=\delta g_{tr}^{(0)}+\cdots\,,\nn
\delta B_t(r)&=\delta B_t^{(0)}+\delta B_t^{(1)}\,r+\cdots,\quad\quad \delta B_r(r)=\frac{\delta B_r^{(0)}}{4\pi T\,r}+\delta B_r^{(1)}+\cdots\,,\nn
\delta B_i(r)&=\delta B_i^{(0)}+\cdots\,,\nn
\delta \phi(r)&=\delta \phi^{(0)}+\cdots,\quad\quad \delta \rho(r)=\delta \rho^{(0)}+\cdots\,.
\end{align}
Moreover, in order to impose regular infalling boundary conditions, the above conditions need to be supplemented by additional constraints,
\begin{align}\label{eq:nh_reg}
\delta g_{tt}^{(0)}+\delta g_{rr}^{(0)}&=2\,\delta g_{rt}^{(0)}\,,\nn
\delta g_{ti}^{(0)}&=\delta g_{ri}^{(0)}\,,\nn
\delta B_r^{(0)}&=\delta B_t^{(0)}\,.
\end{align}

\subsection{Building the hydrodynamic perturbation} \label{sec:build_hydro_mode}

In this subsection, we will focus on the hydrodynamic limit of the bulk theory. This means that we wish to capture the response of the system at small frequencies and long wavelengths compared to e.g. the temperature. However, in the nearly critical system that we consider, there is another small parameter at our disposal, namely the parameter $\ves$ of the previous section, which quantifies the distance from the critical point. As explained in \cite{Donos:2025jxb}, to efficiently describe the coupled dynamics of the order parameter with the hydrodynamic degrees of freedom, we will need to connect all the small parameters available, through the relations $w=\omega\,\ves^2,\,q_i=k_i\,\ves^2$, with $\omega,\,k_i$ of order $\mathcal{O}(\ves^0)$. Similarly to the expansion of the background and of its static fluctuations in powers of $\ves$, we can assume that the radial function $\delta f(r)$ in equation \eqref{eq:fourier_modes} is expandable in $\ves$ as well. In particular, we expand the amplitude fluctuation as,
\begin{align}\label{eq:rho_expeps}
    \delta \rho(r)=\delta \rho_{[1]}(r)\,\ves+\delta\rho_{[3]}(r)\,\ves^3+\mathcal{O}(\ves^5)\,,
\end{align}
and all the rest of the bulk fields as\footnote{The scaling of  the hydrodynamic fluctuations with $\ves$ is dictated by consistency with the equations of motion. Also, the fact that $\delta g_{rt},\,\delta g_{ri},\,\delta B_r$ are zero at order $\mathcal{O}(\ves^0)$ is a (partial) gauge fix of the bulk diffeomorphisms.},
\begin{align}\label{eq:f_expeps}
    \delta f(r)=\delta f_{[2]}(r)\,\ves^2+\delta f_{[4]}(r)\,\ves^4+\mathcal{O}(\ves^6)\,.
\end{align}

By inspecting the expanded equations of motion in $\ves$, we can show that the functions  $\delta \rho_{[1]}(r)$ and $\delta f_{[2]}(r)$ must solve the same set of ODEs with the leading parts of the thermodynamic fluctuations we discussed in Section \ref{sec:Background_linear}. Hence, we can decompose them in the corresponding linear basis of spacetime-independent fluctuations according to,
\begin{align}\label{eq:hydro_exp}
\delta f_{[2]}\,\ves^2=\delta_T f_{[0]}\,\delta T+\delta_\mu f_{[0]}\,\delta\mu+\delta_{v_i}f_{[0]}\,\delta v_i+\delta_{\rho_s} f_{[0]}\,\delta \pi+\delta_{s_{ab}} f_{[0]}\,\delta s_{ab}+\delta_{m_i} f_{[0]}\,\delta m_i\,,\nn
\delta \rho_{[1]}\,\ves=\delta_T \rho_{[0]}\,\delta T+\delta_\mu \rho_{[0]}\,\delta\mu+\delta_{v_i}\rho_{[0]}\,\delta v_i+\delta_{\rho_s} \rho_{[0]}\,\delta \pi+\delta_{s_{ab}} \rho_{[0]}\,\delta s_{ab}+\delta_{m_i} \rho_{[0]}\,\delta m_i\,.
\end{align}
Note that all of the coefficients $\delta T$, $\delta \mu$, $\delta v_i$, $\delta s_{ab}$, $\delta s_a$ and $\delta m_i$ are of order $\mathcal{O}(\ves^2)$, while $\delta \pi$ is of order $\mathcal{O}(\ves^3)$. Moreover, we can expand the gauge invariant combination $\delta m_a$ according to $\delta m_a=\delta m_{a [2]}\ves^2+\delta m_{a[4]}\ves^4+\mathcal{O}(\ves^6)$. 

The coefficients $\delta T,\,\delta \mu,\,\delta v_i$ are naturally interpreted as the temperature, chemical potential and normal fluid velocity fluctuations of the dual field theory. On the other hand, the coefficient $\delta\pi$ is \emph{not} an external source for the amplitude\footnote{ The parameter $\delta\pi$ is essentially the holographic counterpart of $\delta\pi$ in 
  \cite{Donos:2025jxb}, which justifies our notation.}: the corresponding term, $\delta_{\rho_s}\rho _{[0]}\,\delta\pi=\frac{\partial_{\rho_s}\rho_{[0]}}{\ves^2}\,\delta\pi$, entering the expression  \eqref{eq:hydro_exp} for $\delta\rho_{[1]}\,\ves$, is a multiple of the critical mode (see \eqref{eq:mult_rho}) and hence is sourceless.

It is important to stress that the subleading parts of the hydrodynamic fluctuations, $\delta \rho_{[3]}\ves^3$ and $\delta f_{[4]} \ves^4$, are not fixed solely by thermodynamics anymore, but are intimately related to the spacetime dependence of the problem. Specifically, they obey certain inhomogeneous ODEs, in which factors of $\omega$ and $k_i$ appear explicitly.

We choose to turn on an external source for $\delta\rho$ through the asymptotics of its subleading part, namely we take $\delta \rho_{[3]}=\frac{\delta \rho_{[3](s)}}{(r+R)}+\frac{\delta \rho_{[3](v)}}{(r+R)^2}+\cdots,$ and thus, $\delta \rho_{(s)}=\delta \rho_{[3](s)}\ves^3\,.$ The expectation value for the amplitude is,
\begin{align}
\delta\rho_{(v)}=\delta\rho_{[1](v)}\,\ves+\delta\rho_{[3](v)}\,\ves^3+\mathcal{O}(\ves^5)\,.
\end{align}
From equation \eqref{eq:hydro_exp}, we can see that its leading part is,
\begin{align}\label{eq:rho_vev_lead}
\delta\rho_{[1](v)}\,\ves&=\frac{\nu_{T\rho}^{[0]}}{\ves}\,\delta T+\frac{\nu_{\mu\rho}^{[0]}}{\ves}\left(\delta \mu+\mu\,\delta s_{tt}\right)+\frac{\nu_{\rho\rho}^{[0]}}{\ves^2}\delta \pi\Rightarrow\nn
\delta\pi&=\frac{\ves^2}{\nu_{\rho\rho}^{[0]}}\left(\delta\rho_{[1](v)}\,\ves-\frac{\nu_{T\rho}^{[0]}}{\ves}\,\delta T-\frac{\nu_{\mu\rho}^{[0]}}{\ves}\left(\delta \mu+\mu\,\delta s_{tt}\right)\right)\,,
\end{align}
which can be used to trade the parameter $\delta\pi$ for $\delta\rho_{[1](v)}$. We note that based on \eqref{eq:mult_rho} and \eqref{eq:hydro_exp}, it is straightforward to see that $\delta\rho_{[1]}(r)$ is also a scalar multiple of $\delta\rho_{\ast(0)}(r)$,
\begin{align}\label{eq:ampl_mult}
    \frac{\delta\rho_{[1]}(r)}{\delta\rho_{\ast(0)}(r)}=\frac{\delta\rho_{[1]}^{(0)}}{\delta\rho_{\ast(0)}^{(0)}}=\frac{\delta\rho_{[1](v)}}{\delta\rho_{\ast(0)}^v}\,.
\end{align}

By determining the behaviour of the radial component of the complex scalar, we have only fixed the behaviour of the real part of the complex scalar perturbations. To address the imaginary part of the complex scalar, we need to discuss the behaviour of the phase fluctuations in our polar decomposition. For that, we will choose $\delta \theta_{(s)}=\delta\theta_{[2](s)}\,\ves^2$, with $\delta\theta_{[2](s)}=\mathcal{O}(\ves^0)$, determining the complex scalar external source to be $\delta s_\psi=\delta \tilde{s}_\psi\,e^{-i\, w\,t+iq_i x^i}$ with $\delta \tilde{s}_\psi=\delta\rho_{[3] (s)}\ves^3+i q_e\, \rho_v\,\,\delta\theta_{(s)}$ of order $\mathcal{O}(\ves^3)$.

 The next bulk perturbation to which we would like to pay special attention is the one-form $\delta B_\mu$. By examining the expanded equation of motion at leading order in $\ves$, we notice that only radial derivatives of $\delta B_{t\,[2]}\ves^2$ make their appearance. Therefore, we are always free to add a constant of integration $\delta \mu_r\sim \ves^2$ to the perturbation generated by varying the thermodynamic variables of the background. This can be done while imposing infalling boundary conditions on the horizon which also involve the radial component. Moreover, the radial component does not strictly follow \eqref{eq:hydro_exp}, but is fixed, order by order in $\ves^2$ as explained in Appendix \ref{app:delta_B_0}. From the boundary point of view, the asymptotics of equation \eqref{eq:pert_uv_bcs} imply that the physically relevant quantity is the combination,
 \begin{align}\label{eq:mu_r_def}
    \delta \mu+\delta\mu_r=-i\,\omega\,\ves^2\,\delta\theta_{[0](v)}+\delta s_t=\delta m_{t[2]}\ves^2\,.
\end{align}
 
We close this subsection by defining the horizon quantities $\delta g^{(0)}_{{tt}[2n]}$, $\delta g^{(0)}_{{rr}[2n]}$, $\delta B_{r[2n]}^{(0)}$ for positive integer $n$ through the $\ves$-expanded near horizon limits of equation \eqref{eq:gen_exp},
\begin{align}\label{eq:hor_reg_eps}
   \delta g_{tt[2]}(r)&=4 \pi r\, T_c\, \delta g_{tt[2]}^{(0)}+\mathcal{O}(r^2),\quad \delta g_{tt[4]}(r)=4\pi r\left(T_c\, \delta g_{tt[4]}^{(0)}+\frac{\delta T_{\ast(2)}}{2}\delta g_{tt[2]}^{(0}\right)+\mathcal{O}(r^2)\,,\nn
   \delta g_{rr[2]}(r)&=\frac{\delta g^{(0)}_{rr[2]}}{4 \pi T_c\, r}+\mathcal{O}(r^0),\quad \delta g_{rr[4]}(r)=\frac{\delta g^{(0)}_{rr[4]}}{4 \pi T_c\, r}-\frac{\delta g^{(0)}_{rr[2]}}{4 \pi T_c^2\, r}\frac{\delta T_{\ast(2)}}{2}+\mathcal{O}(r^0)\,,\nn
    \delta B_{r[2]}\ves^2&=\frac{\delta B_{r[2]}^{(0)}}{4\pi T_c\,r}+\mathcal{O}(r^0)\,,\quad
    \delta B_{r[4]}=\frac{\delta B_{r[4]}^{(0)}}{4\pi T_c\,r}-\frac{\delta B^{(0)}_{r[2]}}{4 \pi T_c^2\, r}\frac{\delta T_{\ast(2)}}{2}+\mathcal{O}(r^0)\,,
\end{align}
 and similarly for higher $n$. Note that the ``extra'' factors come from substituting the $\ves$ expansion of the temperature, \eqref{eq:eps_intro}, in the near horizon expansions \eqref{eq:gen_exp} and then matching powers of $\ves$ on both sides. The reason is that we must \emph{first} take the near-horizon limit and \emph{then} expand in $\ves$. Moreover, regular ingoing boundary conditions also demand that,
 \begin{align}\label{eq:nh_reg_pert}
\delta g_{tt[2n]}^{(0)}+\delta g_{rr[2n]}^{(0)}=2\,\delta g_{rt[2n]}^{(0)}\,,\quad
\delta g_{ti[2n]}^{(0)}=\delta g_{ri[2n]}^{(0)}\,,\quad
\delta B_{r[2n]}^{(0)}=\delta B_{t[2n]}^{(0)}\,,
\end{align}
for all positive integers $n$.

\subsubsection{Stress Tensor and Current expansion}\label{sec:tj_lead_order}

By Fourier decomposing our bulk modes according to \eqref{eq:fourier_modes}, we are carrying out a similar operation for the boundary quantities. In particular, the stress tensor and $U(1)$ current of the field theory side are Fourier-decomposed as $\delta \langle T^{ab}\rangle=\delta T^{ab}e^{-i\omega \ves^2 t+i \ves^2 k_i x^i}$ and $\delta \langle J^a\rangle=\delta J^a e^{-i\omega \ves^2 t+i \ves^2 k_i x^i}$. For the Fourier coefficients we can similarly write an expansion in $\ves$ according to,
\begin{align}
    \delta T^{ab}&=\delta T^{ab}_{[2]}\,\ves^2+\delta T^{ab}_{[4]}\,\ves^4+\mathcal{O}(\ves^6)\,,\nn
    \delta J^a&=\delta J^{a}_{[2]}\,\ves^2+\delta J^{a}_{[4]}\,\ves^4+\mathcal{O}(\ves^6)\,.
\end{align}
The important observation is that the leading contributions $\delta T^{ab}_{[2]}$ and $\delta J^{a}_{[2]}$ are generated entirely by the thermodynamic perturbations we discussed in Section \ref{sec:Background_linear}. It is straightforward to calculate these  contributions, applying \eqref{eq:stress_curr_uv} and and using the thermodynamic relations \eqref{eq:thermo_rel}, \eqref{eq:trace}, \eqref{eq:phi_susc}\footnote{We raise indices as  $\delta v^i=\eta^{ij}\delta v_j$, $\delta m^i=\eta^{ij}\delta m_j$.},
\begin{align}\label{eq:const_lead}
    \delta T^{tt}_{[2]}\,\ves^2&=\left(c_{\mu[0]} +\mu\,\xi_{[0]}\right) \delta T+\left(\mu\, \chi_{[0]}+T_c\, \xi_{[0]}\right)\left(\delta\mu+\mu\,\frac{\delta \gamma_{tt}}{2}\right)+
     \epsilon_c\, \delta \gamma^{tt}+\left(\mu\,\frac{ \nu_{\mu\rho}^{[0]}}{\ves}+T_c\,\frac{\nu_{T \rho}^{[0]}}{\ves}\right)\delta\pi\,,\nn
    \delta T^{ti}_{[2]}\,\ves^2&=(\epsilon_c+p_c)\delta v^i+\epsilon_c\, \delta \gamma^{it}\,,\nn
    \delta T^{ij}_{[2]}\,\ves^2&=-\,p_c\, \delta \gamma^{ij}+\left(s_c\,\delta T+\varrho_c\,\left(\delta\mu+\mu\,\frac{\delta \gamma_{tt}}{2}\right)\right)\delta^{ij}\,,\nn
    \delta J^t_{[2]}\,\ves^2&=\xi_{[0]}\delta T+\chi_{[0]}\left(\delta\mu+\mu\,\frac{\delta \gamma_{tt}}{2}\right)+\varrho_c\, \frac{\delta \gamma^{tt}}{2}+\frac{\nu_{\mu\rho}^{[0]}}{\ves}\delta\pi\,,\nn
    \delta J^i_{[2]}\,\ves^2&=\varrho_c\,(\delta v^i+\delta \gamma^{it})\,.
\end{align}
Notice that the phase $\delta\theta_{(v)[0]}$ does not appear in \eqref{eq:const_lead}. To write the complete theory at leading order in $\ves$ and close the system of equations, we will need an equation for the time evolution of $\delta\rho_{[1](v)}$. In order to find this equation, as well as the constitutive relations at order $\ves^4$, we will employ the symplectic current technique which we discuss in Section \ref{sec:Apply_SC}.

\subsection{Symplectic current generalities}\label{sec:sympl_curr}

In this subsection, we will briefly review and set the stage for the symplectic current technique, developed in \cite{Donos:2022uea,Donos:2022www,Donos:2022xfd,Donos:2022qao}, to study the hydrodynamic perturbation we are after. The main tool of this method is the Crnkovic-Witten symplectic current defined for any classical theory of a collection of fields $\phi^I$ whose equations of motion can be obtained from a first order Lagrangian density $\mathcal{L}(\phi^I,\partial\phi^I)$. For any two on-shell perturbations $\delta_1\phi^I$ and $\delta_2\phi^I$ around a background $\phi_b^I$, the symplectic current,
\begin{align}\label{eq:scurrent_def}
\mathcal{P}^\mu_{\delta_1,\delta_2}=\delta_1\phi^I\,\delta_2\left(\frac{\partial\mathcal{L}}{\partial( \partial_\mu\phi^I)} \right)-\delta_2\phi^I\,\delta_1\left(\frac{\partial\mathcal{L}}{\partial (\partial_\mu\phi^I)} \right)\,,
\end{align}
is divergence-free,
\begin{align}\label{eq:div_free}
\partial_\mu \mathcal{P}^\mu_{\delta_1,\delta_2}=0\,.
\end{align}
More explicitly, for our holographic setup the contributing terms to \eqref{eq:scurrent_def} read,
\begin{align}\label{eq:sympl_cur_contr}
\frac{\partial\mathcal{L}}{\partial(\partial_\mu g_{\alpha\beta})}&=\sqrt{-g}\,\Gamma^\mu_{\gamma\delta}\left(\,g^{\gamma\alpha}\,g^{\delta\beta}-\frac{1}{2}\,g^{\gamma\delta}g^{\alpha\beta} \right)-\sqrt{-g}\,\Gamma^\kappa_{\kappa\lambda}\,\left(g^{\mu\left(\alpha\right.}g^{\left.\beta\right)\lambda}-\frac{1}{2}\,g^{\mu \lambda}g^{\alpha\beta}\right)\,,\notag\\
\frac{\partial\mathcal{L}}{\partial(\partial_\mu B_\nu)}&=-\sqrt{-g}\,\tau\,F^{\mu\nu}\,,\quad \frac{\partial\mathcal{L}}{\partial(\partial_\mu\phi)}=-\sqrt{-g}\,\partial^\mu\phi\,,\quad \frac{\partial\mathcal{L}}{\partial(\partial_\mu\rho)}=-\sqrt{-g}\,\partial^\mu\rho\,.
\end{align}

When both perturbations $\delta_1$ and $\delta_2$ entering the symplectic current are independent of the dual field theory coordinates, integrating the condition \eqref{eq:div_free} along the radial variable from the horizon to the boundary gives 
\begin{align}\label{eq:constr_deriv}
\mathcal{P}^r_{\delta_1,\delta_2}(r)=\lim_{r\to 0}\mathcal{P}^r_{\delta_1,\delta_2}=\lim_{r\to\infty}\mathcal{P}^r_{\delta_1,\delta_2}\,.
\end{align}
This condition will be used to derive a set of constraints for the thermodynamic variations of Section \ref{sec:Background_linear}, that we list in Appendix \ref{sec:Sympl_constr}.

Let us now take $\delta_1=\delta_H$ to be the hydrodynamic perturbation of the previous subsections and $\delta_2=\delta_{st.}$ to be one of the static perturbations of Section \ref{sec:Background_linear}. Introducing the Fourier modes $P_{\delta_H,\delta_{st.}}^\mu$ for the symplectic current, the condition \eqref{eq:div_free} becomes
\begin{align}\label{eq:sympl_mom}
    {P^r}'_{\delta_H,\delta_{st.}}-i\omega\ves^2 P^t_{\delta_H,\delta_{st.}}-i\omega\ves^2 S' P^r_{\delta_H,\delta_{st.}}+i\ves^2 k_i P^i_{\delta_H,\delta_{st.}}=0\,,
\end{align}
in momentum space. By integrating\footnote{If $\delta \theta_{(s)}\neq0$ there is a small caveat in writing \eqref{eq:sympl_int}, which we address in Appendix \ref{app:diverg}.} over the radial variable from the horizon up to the conformal boundary, we obtain the relation,
\begin{align}\label{eq:sympl_int}
    &\lim_{r\to\infty}P^r_{\delta_H,\delta_{st.}}- \lim_{r\to 0}P^r_{\delta_H,\delta_{st.}}+B_{\delta_H,\delta_{st.}}=0\,,
\end{align}
where we have introduced the bulk integrals
\begin{align}
    \label{eq:Bulk_int_def}B_{\delta_H,\delta_{st.}}=\int^\infty_0 dr \left(-i\omega\ves^2 P^t_{\delta_H,\delta_{st.}}-i\omega\ves^2 S' P^r_{\delta_H,\delta_{st.}}+i\ves^2 k_i P^i_{\delta_H,\delta_{st.}}\right)\,.
\end{align}
Substituting the explicit expressions for $\delta_H$ and the particular $\delta_{st.}$, we can expand the symplectic current in $\ves$ as
\begin{align}\label{eq:eps_sympl}
P^\mu_{\delta_H,\delta_{st.}}={P^\mu}^{[2]}_{\delta_H,\delta_{st.}}\ves^2+{P^\mu}^{[4]}_{\delta_H,\delta_{st.}}\ves^4+\mathcal{O}(\ves^6)\,.
\end{align}

In this work, we are ultimately interested in determining the constitutive relations for the stress tensor and the electric current, as well as
the effective time evolution equation for the amplitude $\delta\rho_{(v)}$ and the phase $\delta\theta_{(v)}$, all up to next-to-leading order in the $\ves$ expansion. To do so, we will need to find the symplectic currents $P^\mu_{\delta_H,\delta_{st.}}$, with\footnote{We will use the static perturbation $\ves\, \delta_{\rho_s}f$ in the symplectic current with the hydrodynamic perturbation, instead of $\delta_{\rho_s}f$, so that the corresponding  symplectic current also obeys \eqref{eq:eps_sympl}.} $\delta_{st.}\in  \{\delta_T,\,\delta_\mu,\,\ves\,\delta_{\rho_s},\,\delta_{v_i},\,\delta_{m_i},\,\delta_{s_{ab}}\}$, up to order $\ves^4$ and examine \eqref{eq:sympl_int} order by order in $\ves$. More specifically, the conformal boundary quantities we will be interested in, can be read off directly from the first term of equation \eqref{eq:sympl_int}. The various $P^{\mu[2n]}_{\delta_H,\delta_{st.}}, n=1,2,\dots$ can be found straightforwardly by applying the definition \eqref{eq:scurrent_def}. We will avoid presenting the explicit expressions for the symplectic current components, since they are rather large and not particularly insightful.

\section{Applying the symplectic current method}\label{sec:Apply_SC}

In this section, we apply the symplectic current technique to our nearly critical holographic system. In Subsection \ref{sec:sc_lead}, we obtain the theory at leading order in $\ves$ where the normal fluid and the charged fluid are coupled through terms which are due to non-dissipative, thermodynamic terms. The next to leading order part of the construction is presented in Subsection \ref{sec:interm_results}. In Subsection \ref{sec:Final_results} we present the full effective theory along with explicit expressions for the transport coefficients in terms of the bulk geometry data.

\subsection{Leading order}\label{sec:sc_lead}

In Subsection \ref{sec:tj_lead_order} we presented preliminary results that fix the leading contribution to the stress tensor and electric current. In this subsection we would like to find the remaining ingredients for a complete theory at leading order in $\ves$. As we have already pointed out, what remains to be fixed is an equation of motion for the amplitude mode $\delta\rho_{[1](v)}$, which, along with the constitutive relations \eqref{eq:const_lead} and the leading order continuity equations following from \eqref{eq:contin}, will form a closed set.

The symplectic current equation \eqref{eq:sympl_int} for $\delta_{st.}=\ves\, \delta\rho_s$ at order $\ves^2$ leads to
\begin{align}\label{eq:ampl_lead}
&\lim_{r\to\infty}P^{r[2]}_{\delta_H,\,\ves\delta_{\rho_s}}-\lim_{r\to 0}P^{r[2]}_{\delta_H,\,\ves\delta_{\rho_s}}=0\Rightarrow\nn
&\frac{\nu_{\mu\rho}^{[0]}}{\ves}\,\delta\mu+\frac{\nu_{T\rho}^{[0]}}{\ves}\,\delta T-\left(T_c\,\frac{\nu_{T\rho}^{[0]}}{\ves}+\frac{\nu_{\rho\phi}^{[0]}}{\ves}\,\phi_s\right)\delta s_{tt}-\delta\rho_{[1](v)}\ves+\frac{\nu_{\rho\rho}^{[0]}}{\ves^2}\,\delta\rho_{[3](s)}\ves^3\nn&+\delta\mu_r\left(\frac{\nu_{\mu\rho}^{[0]}}{\ves}-\frac{\partial_{\rho_s}\varrho_{h[0]}}{\ves}\right)+i\omega\,\ves^2\frac{s_c}{4\pi}\frac{\partial_{\rho_s}\rho^{(0)}_{[0]}}{\ves^2}\delta\rho_{[1]}^{(0)}\ves=0\Rightarrow\nn
&-\delta\pi+\delta\rho_{[3](s)}\ves^3+\frac{\Delta\varrho_{h(2)}\ves^2}{\delta\rho_{\ast (0)}^{v}\ves}\delta\mu_r+i\omega \ves^2\frac{s_c}{4\pi}\frac{\left(\delta\rho_{\ast(0)}^{(0)}\ves\right)^2}{\left(\delta\rho_{\ast(0)}^v\ves\right)^2} \delta\rho_{[1](v)}\ves=0\,.
\end{align}
To obtain the last line, we have used equation \eqref{eq:ampl_mult}, the susceptibility relations \eqref{eq:phi_susc} and the horizon relations \eqref{eq:susc_rho_0}, \eqref{eq:susc_varrho_rel}. We should also stress that by evaluating the symplectic currents $P^\mu_{\delta_H,\delta_\mu}$, $P^\mu_{\delta_H,\delta_T}$ at the same order $\mathcal{O}(\ves^2)$, we get an expression equivalent to \eqref{eq:ampl_lead} and no other extra information. However, to show the equivalence, one must employ the susceptibility relations \eqref{eq:susc_rho_0}, \eqref{eq:susc_varrho_rel}.

The additional information we need is the expression for $\delta\mu_r$, i.e., the Josephson relation. This is found in Appendix \ref{app:delta_B_0}, based on the third of the regularity conditions \eqref{eq:nh_reg}. The result, which we repeat here for convenience, reads,
\begin{align}\label{eq:sol_delta_mur}
    \delta\mu_r=\frac{4\pi}{q_e^2\, s_c\left(\delta\rho_{\ast(0)}^{(0)}\right)^2}\left(i\omega\ves^2\frac{\delta\rho_{[1](v)}}{\delta\rho_{\ast(0)}^v}\Delta\varrho_{h(2)}+q_e^2\left(\delta\rho_{\ast(0)}^v\right)^2\delta\theta_{[2](s)}\ves^2\right)\,.
\end{align}
Notice that $\delta\mu_r$ can be expressed in terms of the phase of the condensate, according to \eqref{eq:mu_r_def}.

In total, the leading-order theory is captured by equations \eqref{eq:const_lead}, \eqref{eq:ampl_lead},\eqref{eq:sol_delta_mur}, and includes the effects of dissipation, due to the presence of the amplitude mode. The fact that viscous effects already appear at leading order is a consequence of connecting the scale of frequency and momenta probing the system with the proximity of the background to criticality (see Section \ref{sec:build_hydro_mode}). On a similar note, as we will see in the next subsections, there are certain non-dissipative terms, which, nonetheless, first contribute to the constitutive relations at next-to-leading order (i.e., at $\mathcal{O}(\ves^4)$).

\subsection{Next-to-leading order}\label{sec:interm_results}

Our goal now is to extract the effective theory at next order in the perturbative expansion. As we have already noted, the first step will be to evaluate the symplectic current relation \eqref{eq:sympl_int} (cf. Appendix \ref{app:diverg}) at order $\ves^4$ for the different choices of thermodynamic perturbations generated by $\delta_{st.}$, as we discussed in Section \ref{sec:Background_linear}. The intermediate steps of this computation are rather bulky to present, but we provide some important expressions in Appendix \eqref{sec:NLO_sympl}. These include constitutive relations for the current (\eqref{eq:Jt_nlo}, \eqref{eq:Ji_nlo}) and stress tensor(\eqref{eq:Ttt_nlo}-\eqref{eq:Tti_nlo_v2}), and an equation of motion for the amplitude mode (\eqref{eq:ampl_nlo}).

Moreover, in the appendix we present the bulk constraint equations \eqref{eq:temp_constr} and \eqref{eq:veloc_constr},  which are particularly useful in obtaining our final results. We have found two distinct expressions for $\delta T^{ti}_{[4]}$ by evaluating $P_{\delta_H,\delta_{s_{ti}}}$ and $P_{\delta_H,\delta_{s_{it}}}$. However, using the bulk constraint \eqref{eq:veloc_constr}, we can check that the two results are actually equivalent. All expressions in \ref{sec:NLO_sympl} contain background data (susceptibilities, horizon data, etc.), the bulk integrals $B_{\delta_H,\delta_{st}}$ (whose explicit expression follows from the definition \eqref{eq:bulk_general}) and the horizon data\footnote{In this respect, the situation is similar to the respective results in \cite{Donos:2022www}; therefore, we shall proceed by following analogous steps.} $\delta g_{tt[4]}^{[0](0)}+\delta g_{rr[4]}^{(0)}$, $\delta g_{ti[4]}^{ (0)}$, $\delta g_{ij[4]}^{(0)}$, $\delta B_{t[4]}^{(0)}$, $\delta B_{i[4]}^{(0)}$, $\delta\rho_{[3]}^{(0)}$, $\delta g_{ti[4]}^{ (1)}$. We can employ \eqref{eq:temp_constr} to eliminate some of them, e.g. $\delta g_{ij[4]}^{(0)}$, in terms of the others. As a final note on this computation, it is useful to mention that the bulk integrals $B_{\delta_H,\delta_{st}}$ simplify considerably after using the background constraints listed in Appendix \ref{sec:Sympl_constr}, along with the continuity equations at leading order.

The next step in our derivation is to fix our hydrodynamic frame. In particular, at this order of the computation in $\ves$, we can perform field redefinitions of the form,
\begin{align}
  \delta \mu\to\delta\mu+\delta\mu_f\,,\quad
  \delta T\to\delta T+\delta T_f\,,\quad
  \delta v^i\to \delta v^i+ \delta v^i_f\,,
\end{align}
with $\delta\mu_f$, $\delta T_f$, $\delta v^i_f$ of order $\mathcal{O}(\ves^4)$, which we can choose to fix the ``transverse frame'' conditions,
\begin{align}\label{eq:t_frame_cond}
    \delta J^t_{diss}=\delta T^{tt}_{diss}=\delta T^{ti}_{diss}=0\,.
\end{align}
In this frame, the components of the stress tensor and electric current with at least one index involving the time coordinate do not receive dissipative corrections as compared to the ideal fluid constitutive relations,
\begin{align}\label{eq:trans_prec}
    \delta \langle J^t\rangle &=\xi\, \delta T+\chi\,\delta \tilde{\mu}+\nu_{\mu\rho}\,\delta \pi+\varrho\, \frac{\delta \gamma^{tt}}{2}+\mathcal{O}(\ves^6)\,,\nn
    \delta \langle T^{tt}\rangle&=\left(c_\mu+\mu\,\xi\right)\delta T+\left(\mu\,\chi+T\,\xi\right)\delta \tilde{\mu}+\epsilon\, \delta \gamma^{tt}+\left(\mu\, \nu_{\mu\rho}+T\,\nu_{T\rho}\right)\,\delta \pi-\rho_v\,\mathrm{Re}(\delta s_\psi)+\mathcal{O}(\ves^6)\,,\nn
    \delta \langle T^{ti}\rangle &=\left(\epsilon+p-\mu^2\chi_{JJ}\right)\delta v^i+\epsilon\,\delta \gamma^{ti}-\mu\,\chi_{JJ}\,\delta m^i+\mathcal{O}(\ves^6)\,.
\end{align}
where we have defined the new hydrodynamic variables,
\begin{align}\label{eq:pi_def}
   \delta \pi=\frac{1}{\nu_{\rho\rho}}\left(\delta\rho_{(v)}-\nu_{T\rho}\,\delta T-\nu_{\mu\rho}\,\delta\tilde{\mu}\right)\,,\quad
   \delta \tilde{\mu}=\delta\mu+\mu\, \frac{\delta \gamma_{tt}}{2}\,.
\end{align}
In the above formulas, it is understood that background quantities carry factors of $\ves$ as well, according to the $\ves$ expansions of Section \ref{sec:Background_linear}\footnote{ This is also true for the susceptibilities appearing in the definition of $\delta \pi$, equation \eqref{eq:pi_def}.}. For instance, the susceptibility $\xi$ in $\delta \langle J^t\rangle$ contributes both at order $\ves^2$ with $\xi_{[0]}\delta T$ and at order $\ves^4$ with $\xi_{[2]}\delta T$. However, the $\xi_{[4]}$ part contributes at order $\ves^6$, i.e. outside the regime of validity of our computation. In addition, we highlight that the non-dissipative terms $-\mu^2 \chi_{JJ}\delta v^i-\mu \chi_{JJ}\delta m^i$ in $\delta \langle T^{ti}\rangle$ first contribute at order $\ves^4$, and thus were absent from the leading-order constitutive relations \eqref{eq:const_lead}.

After following the steps we outlined above, all the bulk integrals and hydrodynamic horizon data drop out of the constitutive relations for the stress tensor and the current. Their precise final form will be presented in the next subsection. However, the resulting equations for the amplitude and the Josephson relation contain the combinations,
\begin{align}
    -i\omega BI_\mu+\delta B_{t[4]}^{(0)}, \qquad -i \omega BI_\rho+i \frac{s_c}{4\pi}\omega\partial_{\rho_s}\rho^{(0)}_{[0]}\delta\rho_{[3]}^{(0)}\,,
\end{align}
respectively. The quantities $BI_\mu$ and $BI_\rho$ are certain bulk integrals of background fields. In order to eliminate the horizon data, $\delta B_{t[4]}^{(0)}$ and $\delta\rho_{[3]}^{(0)}$, we followed the steps sketched in the Appendices \ref{app:delta_B_0} and \ref{app:delta_rho_0}. Notice that these steps  bring in additional bulk integrals which combine with $BI_\mu$ and $BI_\rho$. 

After the manipulations outlined above, a transport coefficient determined by a bulk integral remains both in the amplitude equation and the Josephson relation. This is packaged in the complex transport coefficient $Z_\pi$ introduced in the next subsection. As shown in Appendix \ref{app:Zpi}, $Z_\pi$ characterises the subleading behaviour of the two-point function of the order parameter exactly at the critical point, for zero spatial momentum, and in the zero frequency limit.

\subsection{Final results}\label{sec:Final_results}

We will now summarise the results of the holographic computation in the transverse frame \eqref{eq:t_frame_cond}.
The constitutive relations for the components of the stress tensor and current that were not fixed in equation \eqref{eq:trans_prec}, read,
\begin{align}\label{eq:const_Fin}
\delta \langle T^{ij}\rangle=&-p\, \delta \gamma^{ij}+\left(s\,\delta T+\varrho\,\delta\tilde{\mu}+\rho_v\, \mathrm{Re}(\delta s_\psi)\right)\delta^{ij}- Z_1\, \delta^{ij}\left(\partial_k(\delta v^k+\delta \gamma^{kt})+\partial_t \delta \gamma_{kl}\,\frac{\delta^{kl}}{2}\right)\nn&
    -\eta\,\delta \sigma^{ij}-\rho_v\,\mathrm{Re}\left[Z_2\, \left(\delta\pi-\delta s_\psi\right)\right]\delta^{ij}+\mathcal{O}(\ves^6)\,,\nn
    \delta \langle J^i\rangle=&\left(\varrho-\mu\, \chi_{JJ}\right)\,\delta v^i-\chi_{JJ}\,\delta m^i+\varrho\,\delta \gamma^{ti}-\sigma\,\delta^{ij}\left(\partial_j \delta\tilde{\mu}- \frac{\mu}{T}\,\partial_j\delta T-\delta F_{jt}\right)+\mathcal{O}(\ves^6)\,,
\end{align}
with $\delta\sigma^{ij}$ the fluctuation of the shear tensor,
\begin{align*}
    \delta\sigma^{ij}=2\,\partial^{(i}\delta v^{j)}+2\,\partial^{(i}\delta \gamma^{j)t}+\partial_t\delta \gamma^{ij}-\delta^{ij}\left(\partial_k(\delta v^k+\delta \gamma^{kt})+\partial_t \delta \gamma_{kl}\,\frac{\delta^{kl}}{2}\right)\,.
\end{align*}

We can combine the equation of motion for the amplitude together with the Josephson relation to a complex expression,
\begin{align}
&\partial_t \delta \rho_v+i q_e \rho_v \delta m_t+iq_e\,\mu\,\rho_v\,\frac{\delta \gamma_{tt}}{2}-iq_e\rho_v\, \delta \tilde{\mu}=\nn& \label{eq:ord_linear}-2\, \overline{\Gamma}_0\,\left( -\frac{w_0}{2}\left(\partial_i^2\delta\rho_v+i q_e \rho_v \partial_i(\delta m^i+\mu\, \delta v^i)\right)+\frac{\delta\pi}{2}-\frac{\delta s_\psi}{2}\right)-Z_n\,\rho_v^2 \left(\frac{\delta\pi}{2}-\frac{\delta s_\psi^\ast}{2}\right)\nn&
+Z_2\,\rho_v\left(\partial_k(\delta v^k+\delta \gamma^{kt})+\partial_t \delta \gamma_{kl}\,\frac{\delta^{kl}}{2}\right)-Z_\pi\, \partial_t\left(\frac{\delta\pi}{2}-\frac{\delta s_\psi}{2}\right)+\mathcal{O}(\ves^7)\,,
\end{align}
with $w_0=\chi_{JJ}/(q_e\, \rho_v)^2$. This form of the equation makes it more natural to compare it with the equation of motion for the complex order parameter in field theory constructions \cite{Donos:2025jxb}.

In the formulas above, we have introduced the transport coefficients
$\eta$, $\sigma$, $\Gamma_0$, $Z_1$, $Z_2$, $Z_\pi$ and $Z_n$, which are fixed in terms of the background black hole geometric data. In particular, at leading order in $\ves$, the shear viscosity $\eta$ and the conductivity $\sigma$ are given by,
\begin{align}\label{eq:eta_sigma}
    \eta=\frac{s}{4\pi}+\mathcal{O}(\ves^2)\,,\quad
    \sigma=\tau^{(0)}\frac{s^2 T^2}{(s\, T+\mu\,\varrho)^2}+\mathcal{O}(\ves^2)\,.
\end{align}
However, the expressions for the rest of the transport coefficients simplify considerably if we choose a different thermodynamic ensemble. In particular, up to this point, all susceptibilities of thermodynamic or horizon quantities have been expressed in the $T$, $\mu$, $\rho_s$ ensemble. For the remaining transport coefficients, we found it more practical to present them in the $s,\,\varrho,\,\rho_v$ ensemble. Using the chain rule, it is straightforward to show that the partial derivatives of any thermodynamic quantity $A$ in the two ensembles are related through,
 \begin{align}\label{eq:ch_ens}
     \partial_T (A)_{\mu,\rho_s}&=\frac{c_\mu}{T}\,\partial_s (A)_{\varrho,\rho_v}+\xi\, \partial_\varrho(A)_{s,\rho_v}+\nu_{T\rho}\,\partial_{\rho_v}(A)_{\varrho,s}\,,\nn
     \partial_\mu (A)_{T,\rho_s}&=\xi\,\partial_s (A)_{\varrho,\rho_v}+\chi\, \partial_\varrho(A)_{s,\rho_v}+\nu_{\mu\rho}\,\partial_{\rho_v}(A)_{\varrho,s}\,,\nn
    \partial_{\rho_s} (A)_{\mu,T}&=\nu_{T\rho}\,\partial_s (A)_{\varrho,\rho_v}+\nu_{\mu\rho}\, \partial_\varrho(A)_{s,\rho_v}+\nu_{\rho\rho}\,\partial_{\rho_v}(A)_{\varrho,s}\,.
 \end{align}
For the reader's reference, in Appendix \ref{Other_ens_scale} we give the small $\ves$ behaviour of the various quantities that appear in our transport coefficients in this ensemble.

To simplify the expressions for the remaining transport coefficients $Z_1$, $\Gamma_0$, $Z_n$, $Z_2$ and $Z_\pi$, it is useful to define the quantities,
\begin{align}
     \varpi_1=&\frac{s}{4\pi}\left( \rho^{(0)}\right)^2\,,\quad \varpi_2=\frac{2}{q_e}\,\left(\varrho-\varrho_{h}\right)=\frac{2}{q_e}\Delta\varrho_h\,.
\end{align}
After this definition and using \eqref{eq:ch_ens}, the expressions for $Z_1$, $\Gamma_0$, $Z_n$ and $Z_2$ read,
\begin{align}\label{eq:Transp_final}
    \Gamma_0&=\frac{\rho_v^2}{\varpi_1-i\,\varpi_2}\left(1-i\,\varpi_1 q_e \frac{\partial_\varrho \rho^{(0)}}{\rho^{(0)}}+\partial_\varrho \varrho_h-1\right)\nn
    &\quad+\frac{\rho_v^2}{(\varpi_1-i\,\varpi_2)^2}\left(-i\left(\varpi_2+\rho_v\frac{\partial_{\rho_v}\varrho_h}{q_e}\right)+\varpi_1\left(1-\rho_v \frac{\partial_{\rho_v}\rho^{(0)}}{\rho^{(0)}}\right)\right)\nn
    &\quad-\frac{\rho_v^2}{(\varpi_1-i\,\varpi_2)^2}\frac{s}{8\pi}\left(  i\left(\varpi_1-i\varpi_2\right)q_e\partial_\varrho \phi^{(0)}+\rho_v\partial_{\rho_v}\,\phi^{(0)}\right)^2\,,\nn
    Z_1&=\frac{s}{4\pi}\left(\varrho\, \partial_\varrho \phi^{(0)}+s\,\partial_s\phi^{(0)}\right)^2\,,\nn
    Z_2&=-iq_e\frac{s}{4\pi}\partial_\varrho\phi^{(0)}\left(\varrho\, \partial_\varrho\phi^{(0)}+s\, \partial_s\phi^{(0)}\right)-\frac{i}{(\varpi_1+i\varpi_2)q_e}\left(\frac{\varpi_2}{2}q_e+s\,\partial_s\varrho_h+\varrho\left(\partial_\varrho\varrho_h-1\right)\right)\nn
    &\quad+\frac{s}{4\pi}\frac{\rho_v}{\varpi_1+i\varpi_2}\left(\partial_{\rho_v}\phi^{(0)}\left(\varrho\, \partial_\varrho\phi^{(0)}+s\, \partial_s\phi^{(0)}\right)+\partial_{\rho_v}\rho^{(0)}\left(\varrho\, \partial_\varrho\rho^{(0)}+s\, \partial_s\rho^{(0)}\right)\right)\,,\nn
    Z_n&=\frac{2\,\varpi_1}{\varpi_1^2+\varpi_2^2}\left(1-\partial_\varrho \varrho_h+1-\frac{\rho_v}{\rho^{(0)}}\partial_{\rho_v}\rho^{(0)}-q_e\varpi_2\frac{\partial_{\varrho}\rho^{(0)}}{\rho^{(0)}}\right)\nn&\quad-\frac{s}{4\pi}\frac{1}{\varpi_1^2+\varpi_2^2}\left(q_e^2\,\varpi_1^2  \left(\partial_\varrho\phi^{(0)}\right)^2+\left(\varpi_2 q_e \partial_\varrho\phi^{(0)}+\rho_v \partial_{\rho_v}\phi^{(0)}\right)^2\right)\,.
\end{align}
The above expressions can be trusted up to corrections of order $\mathcal{O}(\ves^2)$, except for the expression for $\Gamma_0$ which is valid up to corrections of order $\mathcal{O}(\ves^4)$. The real and imaginary part of the complex coefficient $Z_\pi$, up to corrections of order $\mathcal{O}(\ves^2$), can be written in terms of background bulk integrals,
\begin{align}\label{eq:ReZ_pi}
\mathrm{Re}Z_\pi=\int^\infty_0dr\left(\frac{2\,\rho_v^2}{e^{2g}\,U\,\rho^2\left(\varpi_1^2+\varpi_2^2\right)^2}\left(\left(\varpi_2^2-\varpi_1^2\right)\left(\tilde{\Phi}_b^2+e^{4g}\rho^4-\varpi_1^2\right)+4\varpi_1^2\varpi_2 \tilde{\Phi}_b\right)-2\right)-2\,R\,,
\end{align}
\begin{align}\label{eq:ImZ_pi}
\mathrm{Im}Z_\pi=\frac{4\,\rho_v^2\, \varpi_1}{(\varpi_1^2+\varpi_2^2)^2} \int^\infty_0dr\frac{1}{e^{2g}\,U\,\rho^2}\left(\left(\tilde{\Phi}_b-\varpi_2\right)\left(\varpi_2\tilde{\Phi}_b+\varpi_1^2\right)+\varpi_2e^{4g}\rho^4\right)\,,
\end{align}
where we defined the radial function,
\begin{align}\label{eq:varpi_defs}
\tilde{\Phi}_b(r)=\frac{2}{q_e}(\Phi_b-\varrho_h)=\frac{\delta \Phi_{b(2)}(r)-\delta\varrho_{\ast h(2)}}{q_e}\,\ves^2+\mathcal{O}(\ves^4)\,.
\end{align}

We will conclude this section with a few comments on our results. The first thing to note is that the form of \eqref{eq:const_Fin}, \eqref{eq:ord_linear} is consistent with the effective theory constructed in \cite{Donos:2025jxb}\footnote{The only difference is in the definition of the chemical potential. In particular, $\delta \mu_{There}=\delta \tilde{\mu}=\delta \mu_{Here}+\mu \frac{\delta \gamma_{tt}}{2}$.}. The explicit expressions \eqref{eq:eta_sigma}, \eqref{eq:Transp_final}, \eqref{eq:ReZ_pi}, and \eqref{eq:ImZ_pi}, for the transport coefficients, teach us that all of them are finite ($\mathcal{O}(\ves^0)$) at the critical point. It is important to observe that all the constraints derived in \cite{Donos:2025jxb}, following from entropy positivity and Onsager's reciprocity, are satisfied by the above expressions for the transport coefficients. For example, the deeper reason that our holographic systems have $\mathrm{Im}Z_n=0$, and that the same coefficient $Z_2$ appears in \eqref{eq:const_Fin} and \eqref{eq:ord_linear}, is Onsager's reciprocity.

The results for the leading part of $\Gamma_0$ and the formulas \eqref{eq:eta_sigma} for the shear viscosity and conductivity are in agreement with previous results for holographic superfluids away from and close to criticality \cite{Donos:2022www},\cite{Donos:2022qao}. However, the $\ves^2$ part of $\Gamma_0$ and the formulas for $Z_1,\,Z_2,\,Z_n,\,Z_\pi$ are novel.

An independent check for the validity of these newly obtained expressions can be carried out by focusing on energy scales much smaller than the gap of the amplitude mode. As we discuss in \cite{Donos:2025jxb}, in this regime, our nearly critical theory leads to first order superfluid hydrodynamics, with the three bulk viscosities of superfluids $\zeta_1,\,\zeta_2,\,\zeta_3$, being fixed in terms of $\Gamma_0, Z_1, Z_2, Z_n$ and background thermodynamic quantities, according to equation (3.6) of \cite{Donos:2025jxb}. Substituting \eqref{eq:Transp_final} in (3.6) of \cite{Donos:2025jxb}, we readily find analytic expressions for $\zeta_i$, which we can trust to first subleading order in $\ves$ (i.e., $\mathcal{O}(\ves^0)$). In addition, in \cite{Donos:2022www}, using the symplectic current, we found analytic formulas for $\zeta_i$ (see equation (4.31) therein), correct to all orders in $\ves$. It is straightforward to expand (4.31) of \cite{Donos:2022www} up to order $\ves^0$ and check that the results agree with the $\zeta_i $ obtained indirectly from \eqref{eq:Transp_final}.

As a final comment, let us consider the case of a scale invariant boundary theory. In that case, the neutral scalar source $\phi_s$ has to be trivial, and as a result the field itself is zero everywhere in the bulk. As we explain in detail in Appendix \ref{app:Conform}, formulas \eqref{eq:Transp_final} then lead to the relations $Z_1=0,\,Z_2=-1$ for the transport coefficients. These results are consistent with the general result of \cite{Donos:2025jxb} for a conformally invariant system, for $d=3$ boundary spacetime dimensions and scaling dimension $\Delta_\psi=2$ for the charged scalar operator.  

\section{Numerical results}\label{sec:Numerics}

In this section, we present numerical checks of our nearly critical effective theory. In particular, we construct numerically all the hydrodynamic quasinormal modes directly from the bulk theory. We compare our numerical results with the predictions of our effective theory close to the critical point.

\subsection{Setup and description of the numerical method}

The bulk action we choose for our numerical computation is given by \eqref{eq:bulk_action} with the specific choices,
\begin{align}
    V&=-6+\frac{1}{2}m_\psi^2\, |\psi|^2+\frac{1}{2}m_\phi^2\,\phi^2+\lambda_\rho |\psi|^4+\lambda_\phi \phi^4+\lambda_{\rho\phi}|\psi|^2 \phi^2\,,\nn\,
    \tau&=1+\zeta_\rho\, |\psi|^2\,.
\end{align}
A specific parameter choice, compatible with our general requirements of Section \ref{sec:setup} is,
\begin{align}
m_\psi^2&=m_\phi^2=-2,\quad q_e=1,\nn
\lambda_\rho&=\lambda_\phi=\frac{1}{2},\quad \lambda_{\rho\phi}=-1,\quad \zeta_\rho=1\,.
\end{align}

In order to construct our background black hole solutions, we solve the equations of motion \eqref{eq:eom} after substituting the ansatz \eqref{eq:background}. The boundary conditions we need to impose to our system of ODEs are given by equation \eqref{eq:nh_bexp} near the black hole horizon and \eqref{eq:nb_exp} near the asymptotically AdS  boundary. Given that we wish to study spontaneous symmetry breaking, we also fix $\rho_s=0$ as a boundary condition. This procedure results in a 3-parameter family of numerical background solutions, labelled by the temperature $T$, the chemical potential $\mu$ and the neutral scalar source $\phi_s$. 

By choosing $\phi_s/\mu=1$ we find that the system undergoes a second order phase transition at a critical temperature $T_c/\mu= 0.1278\ldots$. In practice, this means that for $T>T_c$ we find a unique solution with trivial complex scalar, corresponding to the normal phase of our system. For $T<T_c$, we find an additional solution with non-trivial charged scalar, corresponding to the superfluid phase of our system.

After having constructed the background black holes, our ultimate goal is to study numerically spacetime-dependent linear fluctuations of the form,
\begin{align}\label{eq:time_dep_pert}
\delta\mathcal{F}(t,x;r)=e^{-i\, \omega\,(t+S_n(r))+iq_1 x}\,\delta f(r)\,.
\end{align}
In particular, we would like to examine the limit of their boundary characteristics close to criticality. Given the spacetime translational symmetry of the background \eqref{eq:background}, we have Fourier transformed the fluctuation. Isotropy allows us to align the wavevector with the $x$ axis without loss of generality.

The function $S_n(r)$ serves a similar purpose as the function $S(r)$ that we introduced in equation \eqref{eq:fourier_modes}. We will choose $S_n(r)=\int^r_\infty dr' \frac{1}{U(r')}$, which is an appropriate choice for infalling boundary conditions on the horizon, provided that the radial function $\delta f(r)$ is analytic there. This is exactly the condition we imposed in Subsection \ref{sec:space_time_perts}. However, the falloff of the specific choice for $S_n(r)$ close to the boundary is $\sim 1/r$. This suggests that extra care is needed when extracting field theory quantities from the asymptotics of our perturbations.

As we would expect, by examining the equations of motion after substituting the ansatz \eqref{eq:time_dep_pert} for the perturbation, we can deduce that the fields split into two decoupled sectors. The longitudinal sector, involving the fluctuations $\{ \delta g_{tt}$, $\delta g_{tr}$, $\delta g_{tx}$, $\delta g_{rr}$, $\delta g_{rx}$, $\delta g_{xx}$, $\delta g_{yy}$, $\delta b_t$, $\delta b_r$, $\delta b_x$, $\delta\rho$, $\delta\phi \}$ and the transverse sector, which involves $\{\delta g_{ty}$, $\delta g_{xy}$, $\delta g_{ry}$, $\delta b_y\}$.

Focusing on the longitudinal sector, we note that the $r$ component gauge field equation is algebraic for $\delta b_r$. We can  thus use it to eliminate $\delta b_r$ from the rest of the equations. Next, we fix the diffeomorphism invariance by choosing the radial gauge with,
\begin{align}\label{eq:radial_gauge}
    \delta g_{tr}(r)=\delta g_{rr}(r)=\delta g_{rx}(r)=0\,.
\end{align}
After eliminating $\delta b_r$, we have eleven ODEs in the longitudinal sector. However, as a consequence of the Bianchi identities, we can check that not all of the differential equations are independent. More specifically, we can show that we have in fact eight independent ODEs. A first order ODE for each one of $\delta g_{tt},\delta g_{xx},\delta g_{tx}$ and a second order ODE for $\delta g_{yy},\delta b_t,\delta b_x,\delta\rho,\delta\phi$. One can check that the remaining three ODEs that we left out of our system are automatically satisfied.

With the final equations at hand, it will be useful to examine the asymptotics of our fields close to the horizon at $r\to 0$ and at infinity, close to the boundary. Close to the horizon, the infalling condition suggests the expansions,
\begin{align}
    \delta g_{tt}&=\mathcal{O}(r^2),\quad \delta g_{tx}=\mathcal{O}(r),\quad  \delta g_{xx}=\delta g_{xx}^{(0)}+\mathcal{O}(r),\quad \delta g_{yy}=-\delta g_{xx}^{(0)}+\mathcal{O}(r),\nn
    \delta b_t&=\delta b_t^{(0)}+\mathcal{O}(r), \quad  \delta b_x=\delta b_x^{(0)}+\mathcal{O}(r),\quad \delta \rho=\delta \rho^{(0)}+\mathcal{O}(r),\quad \delta\phi=\delta \phi^{(0)}+\mathcal{O}(r).
\end{align}
The coefficients of all the terms that do not appear explicitly above are not independent but are fixed by the five constants of integration $\delta g_{xx}^{(0)}$, $\delta b_t^{(0)}$, $\delta b_x^{(0)}$, $\delta \rho^{(0)}$ and $\delta \phi^{(0)}$ by expanding the equations of motion in the radial coordinate. Near the conformal boundary, we find that our fields follow the expansions,
\begin{align}\label{eq:num_bound_exp}
   \delta g_{tt}&=\delta g_{tt}^s(r+R)^2+\cdots+\frac{\delta g_{tt}^v}{r+R}+\cdots,\nn
   \delta g_{xx}&=\delta g_{xx}^s(r+R)^2+\cdots,\quad
   \delta g_{tx}=\delta g_{tx}^s(r+R)^2+\cdots,\quad
   \delta g_{yy}=\delta g_{yy}^s(r+R)^2+\cdots,\nn
   \delta b_t&=\delta b_t^s+\frac{\delta b_t^v}{r+R}+\cdots,\quad
   \delta b_x=\delta b_x^s+\cdots,\nn
   \delta \rho&=\frac{\delta \rho_s}{r+R}+\frac{\delta \rho_v}{(r+R)^2}+\cdots,\quad
    \delta \phi=\frac{\delta \phi_s}{r+R}+\frac{\delta \phi_v}{(r+R)^2}+\cdots.
\end{align}
The coefficients of all the terms that do not appear explicitly above are not independent but are fixed by the twelve constants of integration $\delta g_{tt}^s$, $\delta g_{tt}^v$, $\delta g_{xx}^s$, $\delta g_{tx}^s$ , $\delta g_{yy}^s$, $\delta b_t^s$, $\delta b_t^v$, $\delta b_x^s$, $\delta \rho_s$, $\delta \rho_v$, $\delta \phi_s$ and $\delta \phi_v$.

We wish to solve our system of ODEs using a double sided shooting method. However, before doing so, we need to understand better a residual gauge freedom close to the AdS boundary. More specifically, we can search for a diffeomorphism $x^\mu\to x^\mu+\hat{\xi}^\mu(r)e^{-i\omega t+i q_1\, x}$, $\mu=t,r,x$ that satisfies the conditions \eqref{eq:radial_gauge} near the boundary. This suggests that near the conformal boundary,
\begin{align}
    \hat{\xi}^r&=(r+R) c_r+\mathcal{O}\left(\frac{1}{r+R
}\right)\,,\quad
    \hat{\xi}^t=c_t+\frac{i\,\omega\,  c_r}{2}\frac{1}{(r+R)^2}+\cdots\,,\quad
    \hat{\xi}^x=c_x+\frac{i\,q_1\,  c_r}{2}\frac{1}{(r+R)^2}+\cdots\,.
\end{align}
Adding the pure-gauge fields that result from such a coordinate transformation to our perturbation is equivalent to shifting the constants of integration appearing in \eqref{eq:num_bound_exp} according to,
\begin{align}
    &\delta g_{tt}^s\to\delta g_{tt}^s+2\,c_r-2i\omega\, c_t\,,\quad \delta g_{xx}^s\to \delta g_{xx}^s-2\,c_r-2\,iq_1\, c_x\,,\quad \delta g_{tx}^s\to \delta g_{tx}^s+i\omega c_x+iq_1c_t\,,\nn
    &\delta g_{yy}^s\to \delta g_{yy}^s-2\,c_r\,,\quad \delta b_t^s\to \delta b_t^s+i\omega c_t\,\mu\,,\quad \delta b_x^s\to \delta b_x^s-iq_1 c_t\,\mu\,,\quad \delta\phi_s \to \delta\phi_s+c_r\phi_s\,,\nn
    &\delta g_{tt}^v\to
    \delta g_{tt}^v+\frac{i\omega\, c_r}{12}\left(\phi_s^2-8\, \omega^2\right)+\frac{\omega\, c_t}{3}\left(-6\, i\, g_{(v)}+\omega\, \phi_s^2+\omega^3\right)\,,\nn
    & \delta b_t^v \to \delta b_t^v-i\omega\varrho\, c_t+\varrho\, c_r\,,\quad
    \delta\phi^v\to \delta\phi^v+2c_r\,\phi_v\,,\quad \delta\rho^v\to \delta\rho^v+2c_r\,\rho_v\,.
\end{align}
Similarly, there is a residual gauge freedom pertinent to the $U(1)$ gauge symmetry in the bulk. More precisely, a pure-gauge configuration $\delta b_t=-i\omega \Lambda(r)\,,\delta b_x=iq_1\Lambda(r)$  also solves the equations of motion close to the boundary, provided $\Lambda$ asymptotes the behaviour,
\begin{align}
\Lambda(r)=\Lambda_0+\mathcal{O}\left(\frac{1}{(r+R)^2}\right)\,,
\end{align}
close to the conformal boundary. Adding such a configuration to our perturbation, effectively shifts the leading terms of $\delta b_t\,,\delta b_x$ in \eqref{eq:num_bound_exp} according to,
\begin{align}
    \delta b_t^s\to \delta b_t^s+i\omega c_t\,\mu-i\omega\Lambda_0\,,\quad \delta b_x^s\to \delta b_x^s-iq_1 c_t\,\mu+iq_1\Lambda_0\,.
\end{align}

To summarise, taking into account the extra constants coming from $\xi^\mu$ and $\Lambda$ we have in total twenty one integration constants. As we have mentioned, we have a coupled system of three first and five second order ODEs to solve, amounting to 13 matching conditions. If we are after a Green's function, apart from fixing $\omega$ and $q_1$ this means that we have 8 free constants to fix, which are exactly the 8 sources  $(\delta g_{tt}^s$, $\delta g_{xx}^s$, $\delta g_{tx}^s$, $\delta g_{yy}^s$, $\delta b_t^s$, $\delta b_x^s,$, $\delta \rho_s$, $\delta \phi_s)$ of the expansion \eqref{eq:pert_uv_bcs} that we actually want to hold fixed in our shooting method. On the other hand, if we are looking to calculate a quasinormal mode, then after picking $q_1$  and setting all the sources to zero, we have an extra free constant, since the frequency is to be found. This constant can be any of the remaining integration constants and can be set to 1, due to the linearity and homogeneity of the ODEs.

We will now briefly discuss the transverse sector, which is very similar to the case of the longitudinal one. To begin, we fix the coordinate invariance by picking the radial gauge $\delta g_{ry}=0$. There are four ODEs in the transverse sector but only three of them are independent. These can be taken to be a first order ODE for $\delta g_{xy}$ and a second order ODE for $\delta g_{ty}$ and $\delta b_y$. The fourth ODE of the original system is then a simple consequence of the reduced one.

Following similar steps, close to the horizon we have the transverse sector expansions,
\begin{align}
    \delta g_{ty}=\mathcal{O}(r),\,\delta g_{xy}=\delta g_{xy}^{(0)}+\mathcal{O}(r),\,\delta b_y=\delta b_y^{(0)}+\mathcal{O}(r)\,.
\end{align}
Once again, the coefficients of all the terms that do not appear explicitly above are not independent but are fixed by the two constants of integration $\delta g_{xy}^{(0)}$ and $\delta b_y^{(0)}$. Close to the conformal boundary, the fields asymptote to,
\begin{align}
    \delta g_{ty}&=(r+R)^2\delta g_{ty}^s+\cdots+\frac{\delta g_{ty}^v}{r+R}+\cdots,\, \delta g_{xy}=(r+R)^2\delta g_{xy}^s+\cdots\,,\nn
    \delta b_y&=\delta b_y^s+\frac{\delta b_y^v}{r+R}+\cdots\,,
\end{align}
with the the remaining coefficients fixed by the five constants $\delta g_{ty}^s$, $\delta g_{ty}^v$, $\delta g_{xy}^s$, $\delta b_y^s$ and $\delta b_y^v$.

In the transverse sector, the residual diffeomorphism invariance is given by the transformation $y \to y+\hat{\xi}^y(r)e^{-i\omega t+i q_1 x}$, which keeps us in the radial gauge provided that $\hat{\xi}^y=c_y+\cdots$ near the boundary. Such an operation would shift our integration constants according to,
\begin{align}
\delta g_{ty}^s\to \delta g_{ty}^s+i\omega\, c_y\,,\quad\delta g_{xy}^s\to \delta g_{xy}^s-iq_1\,c_y\,,\quad \delta g_{ty}^v \to \delta g_{ty}^v-\frac{i\omega}{3}c_y\phi_s\phi_v\,.
\end{align}
Taking into account the extra constant $c_y$ coming from $\xi^y$, we have a total of eight integration constants. At the same time, we have five matching conditions coming from the double sided shooting method.

To calculate a Green's function in this sector, apart from fixing $\omega , q_1$ we have three free parameters, which are the three sources $\delta g_{ty}^s,\,\delta g_{xy}^s,\, \delta b_y^s$. To calculate the quasinormal modes, we set the three sources to zero and one of the other integration constants equal to 1 (invoking the linearity and homogeneity of the ODEs once more).

\subsection{Quasinormal modes}

In this subsection, we will numerically construct the hydrodynamic quasinormal modes of the bulk theory and compare their dispersion relations to those of the effective theory we presented in Subsection \ref{sec:Final_results}.

To find the dispersion relations of the quasinormal modes from our effective theory, we first set all external sources to zero. The continuity equations for the stress tensor and the $U(1)$ current \eqref{eq:contin}, along with the complex equation of motion for the order parameter \eqref{eq:ord_linear} form a $6\times 6$ linear homogeneous system of equations for the 6 unknown Fourier mode components $\delta T$, $\delta\mu$, $\delta\theta_{(v)}$, $\delta v^i$ and $\delta\rho_{v}$. In order to obtain a nontrivial solution, the matrix of coefficients of the linear system has to be non-invertible. Setting its determinant to zero, we obtain a sixth order polynomial equation for $\omega$, whose roots determine the dispersion relations $\omega(\vec{q})$.

The six solutions of this equation correspond to the six hydrodynamic modes of the theory. From the viewpoint of superfluids away from criticality, these account for two\footnote{Parity invariance of the thermal background state implies that if $\omega(q_1,q_2)$ is a pole of  a Green's function, so is $\omega(-q_1,q_2)$.} first sound modes $\omega_{FS,\pm}(\vec{q})$, two second sound/Goldstone modes $\omega_{SS,\pm}(\vec{q})$, one Higgs/amplitude mode $\omega_{H}(\vec{q})$ and one shear diffusive mode $\omega_{D}(\vec{q})$ describing diffusion of spatial momentum to a transverse direction.

\begin{figure}[h!]
\centering
\includegraphics[width=0.49\linewidth]{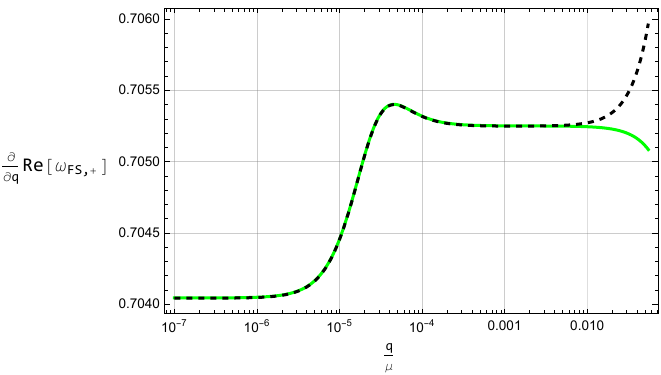}\hspace{0.2em}\includegraphics[width=0.49\linewidth]{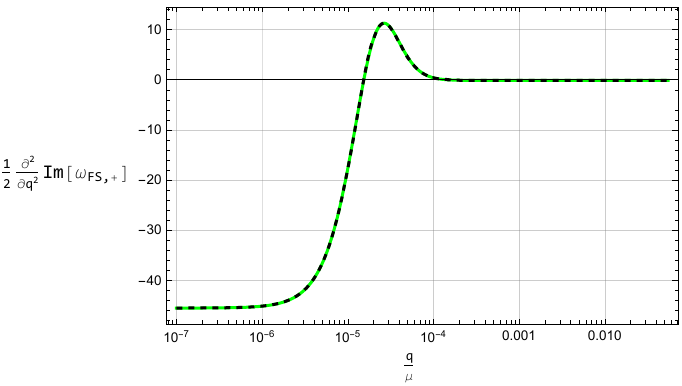}
\caption{Plots of $\frac{\partial}{\partial q}\mathrm{Re}[\omega_{FS,+}]$ and $\frac{1}{2}\frac{\partial^2}{\partial q^2}\mathrm{Im}[\omega_{FS,+}]$ for the first sound mode as a function of $\frac{q}{\mu}$ for  $\frac{T}{T_c}\approx0.9999922$. The dashed lines correspond to the numerical results and the solid green lines to the analytic predictions.}
\label{fig:first_sound}
\end{figure}

We have numerically constructed these modes in the bulk for several values of the wavevector $q$, at fixed temperature\footnote{The next-to-leading order effective theory we have found analytically in the previous sections is valid for a range of temperatures close to the critical point. We choose a temperature this close to $T_c$ to perform our numerics only because the interpolating behaviour of the quasinormal modes is more pronounced then, as can be clearly seen in the figures below.} $\frac{T}{T_c}\approx0.9999922$, corresponding to $\ves^2\approx \frac{T-T_c}{T_c}\approx 10^{-5}$. Figure \ref{fig:first_sound} shows $\frac{\partial}{\partial q}\mathrm{Re}[\omega_{FS,+}]$ and $\frac{1}{2}\frac{\partial^2}{\partial q^2}\mathrm{Im}[\omega_{FS,+}]$ for the first sound mode. As is evident from the plots, this mode interpolates between the following asymptotic behaviours: For small $\frac{q}{\mu}$, it is the first sound of the superfluid phase, with $\omega\simeq 0.70404\,  q-i\,45.5\,q^2$ and for large $\frac{q}{\mu}$ it behaves like the first sound of the normal phase, with $\omega\simeq 0.70525\,  q-i\,0.1602\,q^2$.  Notice that for small momenta the coefficient of the $q^2$ term, i.e. the attenuation constant, is significantly large (compared, e.g. to the corresponding speed of sound), in accordance with our remark in Section 3.1 of \cite{Donos:2025jxb} that it diverges close to the critical point.

\begin{figure}[h!]
\centering
\includegraphics[width=0.48\linewidth]{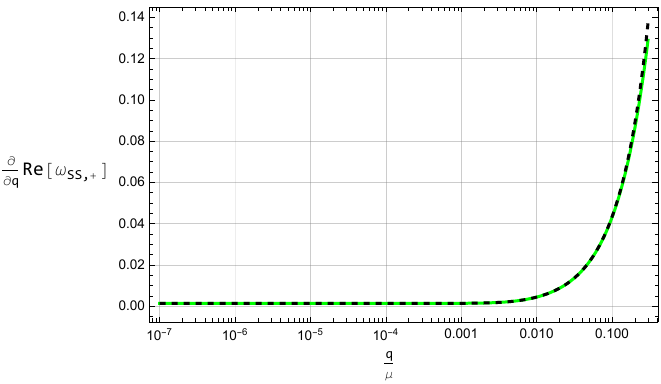}\quad\includegraphics[width=0.48\linewidth]{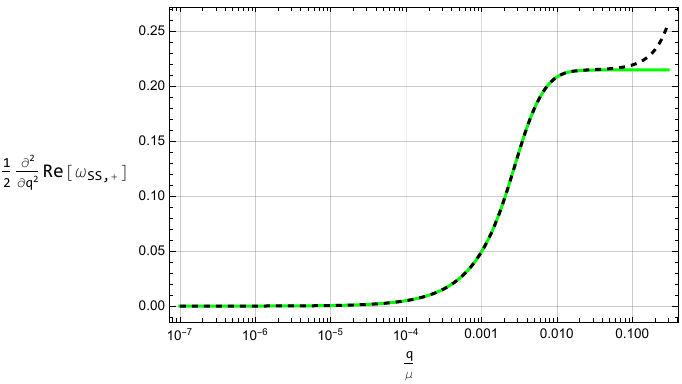}\\
\includegraphics[width=0.48\linewidth]{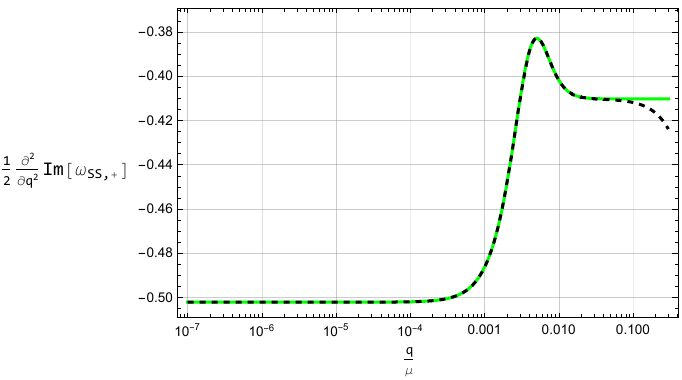}
\caption{Plots of $\frac{\partial}{\partial q}\mathrm{Re}[\omega_{SS,+}]$,  $\frac{1}{2}\frac{\partial^2}{\partial q^2}\mathrm{Re}[\omega_{SS,+}]$ and $\frac{1}{2}\frac{\partial^2}{\partial q^2}\mathrm{Im}[\omega_{SS,+}]$ for the second sound mode as a function of $\frac{q}{\mu}$ for  $\frac{T}{T_c}\approx0.9999922$. The dashed lines correspond to the numerical results and the solid green lines to the analytic predictions.}
\label{fig:second_sound}
\end{figure}

In Figure \ref{fig:second_sound}, we plot $\frac{\partial}{\partial q}\mathrm{Re}[\omega_{SS,+}]$,  $\frac{1}{2}\frac{\partial^2}{\partial q^2}\mathrm{Re}[\omega_{SS,+}]$ and $\frac{1}{2}\frac{\partial^2}{\partial q^2}\mathrm{Im}[\omega_{SS,+}]$ for the second sound mode. As we can see from the plots, for small $\frac{q}{\mu}$ this mode behaves as $\omega \simeq 0.0012968\,q-i\,0.5021\, q^2$, being the Goldstone mode of the superfluid phase. Notice that the speed of the Goldstone mode is significantly small, which is expected, since it is proportional\cite{Donos:2022www} to $\sqrt{\chi_{JJ}}\sim\ves$, close to the critical point. Moreover, we highlight that the attenuation constant for the Goldstone mode is $0.5021\sim\mathcal{O}(\ves^0)$ justifying our comment in Section 3.1 of \cite{Donos:2025jxb}, that it remains finite close to the transition. For large values of $\frac{q}{\mu}$, this mode behaves as $\omega \simeq (0.215-0.41\,i)q^2$, matching the mode $\omega_{r,1}=-iw_0\,\Gamma_0\,k^2$, which is one of the two modes of the order parameter in the normal phase close to the critical point, as explained in Section 3.2 of \cite{Donos:2025jxb}.

\begin{figure}[!h]
\centering
\includegraphics[width=0.49\linewidth]{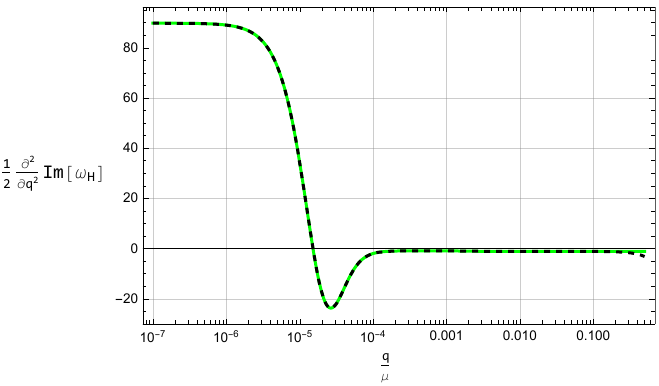}\hspace{0.2em}\includegraphics[width=0.49\linewidth]{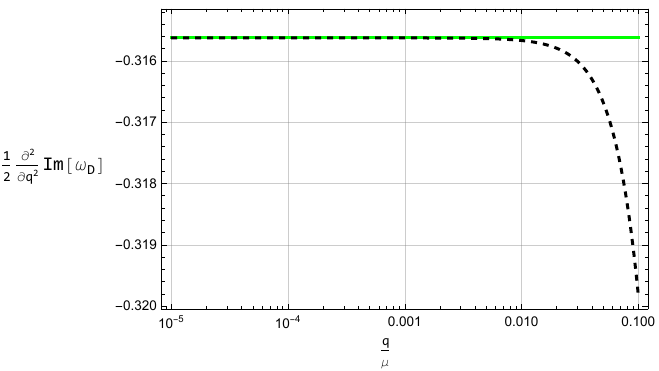}
\caption{Plots of $\frac{1}{2}\frac{\partial^2}{\partial q^2} \mathrm{Im}[\omega_H]$ and $\frac{1}{2}\frac{\partial^2}{\partial q^2} \mathrm{Im}[\omega_D]$, for the Higgs and shear diffusive mode respectively,  as a function of $\frac{q}{\mu}$, for $\frac{T}{T_c}\approx0.9999922$. The dashed lines correspond to the numerical results and the solid green lines to the analytic predictions.}
\label{fig:higgs_shear}
\end{figure}

Figure \ref{fig:higgs_shear} depicts $\frac{1}{2}\frac{\partial^2}{\partial q^2} \mathrm{Im}[\omega_H]$ and $\frac{1}{2}\frac{\partial^2}{\partial q^2} \mathrm{Im}[\omega_D]$, for the Higgs and shear diffusive mode respectively. The Higgs mode for small values of the ratio $\frac{q}{\mu}$ approaches $\omega_H\simeq -0.00001874\,i+i\,89.7 q^2$, in accordance with equations 3.8-3.10 of \cite{Donos:2025jxb}. The diffusion constant $D_H$ is indeed negative and significantly large, which agrees with the diverging behaviour predicted in Section 3.1 of \cite{Donos:2025jxb}. We should stress that despite the negative value of $D_H$, the $\omega_H$ mode remains in the lower half of the complex $\omega$ plane for all values of the wavevector $q$. For large values of $\frac{q}{\mu}$, the Higgs mode behaves as $\omega \simeq -1.1 \,i\,q^2$, matching the dispersion relation of the charge diffusion mode of the normal phase, as explained in Section 3.2 of \cite{Donos:2025jxb} . Finally, the shear diffusive mode has $\omega\simeq-0.315629 \,i\,q^2$ without showing any interpolating behaviour, as the other modes do. This is expected since its diffusion constant is given by $\frac{\eta}{s T+\mu\varrho}$ (see for example  \cite{Donos:2022www}), which is continuous across the transition.

Overall, we find very good quantitative agreement between the analytic predictions and the numerical results. We emphasize that this agreement holds for the entire interpolating behaviour of the modes, as well as the asymptotic regions of small and large $\frac{q}{\mu}$ compared to $\frac{T-T_c}{T_c}$. The plateaus of the plots for large wavenumbers, corresponding to the normal phase modes, are expected to grow larger and more pronounced as we keep approaching the critical temperature. On the other hand, moving far away from the critical point these plateaus would gradually shrink to zero, and the only hydrodynamic behaviour visible would be the one of the superfluid broken phase, emerging from the regime of small wavenumbers.

We would like now to focus on the gap of the Higgs mode within our effective theory. As shown in \cite{Donos:2025jxb}, the gap is given by,\footnote{Where,  $\tilde{\nu}_{\rho\rho}\equiv\left(\frac{\partial \rho_v}{\partial \rho_s}\right)_{s,\varrho}$.}
\begin{align}\label{eq:Gap_new}
    \omega_g=-i\frac{2\,\mathrm{Re}\Gamma_0+\mathrm{Re}Z_n\, \rho_v^2}{2\,\tilde{\nu}_{\rho\rho}+\mathrm{Re}Z_\pi}+\mathcal{O}(\ves^6)\,.
\end{align} 
We note that this formula captures both the leading ($\sim\ves^2$) and the next-to-leading ($\sim \ves^4$) behaviour of the gap. Hence, it improves and extends the validity of the corresponding formula found in \cite{Donos:2022xfd} (and later in \cite{Donos:2022qao}),
\begin{align}\label{eq:Gap_old}
    \omega_g=-i\frac{\,\mathrm{Re}\Gamma_0^{lead.}}{\,\tilde{\nu}_{\rho\rho}}+\mathcal{O}(\ves^4)\,,
\end{align}
with,
\begin{align}
    \Gamma_0^{lead.}=\frac{\rho_v^2}{\varpi_1-i \varpi_2}\,,
\end{align}
which is correct only to leading order in $\ves$. 

\begin{figure}[!h]
\centering
\includegraphics{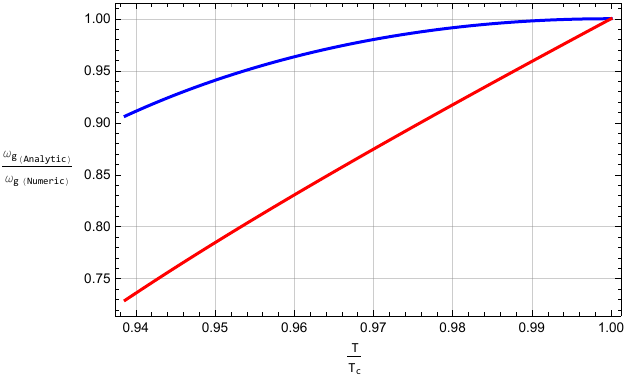}
\caption{Plot of $\frac{\omega_{g(Analytic)}}{\omega_{g (Numeric)}}$ for a range of temperatures close to the critical one. The gap $\omega_{g (Numeric)}$ was calculated numerically. The gap $\omega_{g(Analytic)}$ of the blue line was calculated with \eqref{eq:Gap_new}, and $\omega_{g(Analytic)}$ of the red line was calculated using \eqref{eq:Gap_old}.}
\label{fig:higgs_Gap}
\end{figure}

In Figure \ref{fig:higgs_Gap} we plot the ratio $\frac{\omega_{g(Analytic)}}{\omega_{g (Numeric)}}$ for a range of temperatures close to the critical one. The Higgs gap $\omega_{g (Numeric)}$ was calculated numerically. For the blue line, the gap $\omega_{g(Analytic)}$  was calculated using the new formula \eqref{eq:Gap_new}, while for the red line, the gap $\omega_{g(Analytic)}$  was calculated using the old formula \eqref{eq:Gap_old}. We clearly see in the diagram that the new formula approximates the gap with much better accuracy than the old one, as we move away from $T_c$, as expected. Since the coefficient $Z_\pi$ features in the expression \eqref{eq:Gap_new} for the gap, this numerical check validates the quantitative importance of this novel coefficient in the accurate description of the low energy spectrum.

\section{Discussion}\label{sec:Discuss}

We have studied the perturbative dynamics of holographic superfluids in the vicinity of the second order phase transition. We moved away from the probe limit of our previous work \cite{Donos:2022qao}, taking into account fluctuations of the metric tensor as well. From a field theory perspective, this corresponds to tackling the complete dynamics of the complex order parameter of superfluids, coupled to chemical potential, temperature, and normal fluid velocity fluctuations.

By employing the symplectic current method of \cite{Donos:2022uea,Donos:2022www,Donos:2022xfd,Donos:2022qao}, we derived the constitutive relations for the stress tensor and the electric current, as well as an equation governing the order parameter, at next-to-leading order in a certain perturbative expansion scheme. Moreover, we have obtained analytic formulas for all the transport coefficients in terms of background quantities. 

Most importantly, our findings provide an independent verification of the validity of the effective theory we proposed in \cite{Donos:2025jxb}. This effective description is a covariant generalisation of \cite{KhalatnikovLebedev1978}, but crucially involves an additional term in the order parameter equation, with complex coefficient $Z_\pi$.

A possible future direction could be to apply the symplectic current technique to other interesting systems and derive their hydrodynamic behaviour. A prominent example would be the study of hydrodynamics of holographic systems with broken translations \cite{Blake:2015epa, Gouteraux:2016wxj }. In particular, the effective theory for their nearly critical regime has not been fully studied yet, even outside holography.

Another interesting question regards the symplectic current method itself. In principle, the symplectic current allows us to study the \emph{linearised} hydrodynamic behaviour of any holographic system analytically, without having to tackle the equations of motion in the bulk. It would be interesting to investigate whether the technique could be generalised in order to study nonlinear effects in the small-frequency, large-wavelength regime of holographic systems.

\section*{Acknowledgements}

AD and PK are supported by the Leverhulme Research Project Grant RPG-2023-058. AD is supported by STFC grant ST/T000708/1.

\appendix

\section{Susceptibility relations}\label{app:susc_rel}

Equation \eqref{eq:mult_rho}, i.e. the observation that the amplitude of a static perturbation $\delta_{st.}\rho$ for $\delta_{st.}\in\{\delta_T\,,\delta_\mu\,,\delta_{\rho_s}\}$ is, to leading order in $\ves$, a scalar multiple of the critical mode $\delta \rho_{\ast(0)}$ leads to several useful relations which we gather in this appendix.

To obtain the first of these, we simply take the near horizon limit in \eqref{eq:mult_rho} to find
\begin{align}\label{eq:susc_rho_0}
\frac{\partial_T\rho^{(0)}_{[0]}}{\nu_{T\rho}^{[0]}}=\frac{\partial_\mu\rho^{(0)}_{[0]}}{\nu_{\mu\rho}^{[0]}}=\frac{\partial_{\rho_s}\rho^{(0)}_{[0]}}{\nu_{\rho\rho}^{[0]}}=\frac{\delta\rho_{\ast (0)}^{(0)}}{\delta\rho_{\ast (0)}^{v}}\,.
\end{align}

To obtain the rest, we consider the bulk charge of a static perturbation $\delta_{st.}$,  which, upon linearising \eqref{eq:bulk_charge_1}, obeys
\begin{align}
 \delta_{st.} \Phi'(r)=q_e^2\, \delta_{st.}\left(\frac{e^{2g}}{U}\,a\,\rho^2\right)\,.
\end{align}
In addition, expanding it in $\ves$, we have
\begin{align}
    \delta_{st.} \Phi(r)=\frac{1}{\ves^{n_{st.}}}\left(\partial_{st.} \Phi_{[0]}(r)+\mathcal{O}(\ves^2)\right)\,,
\end{align}
with $n_{st.}=0$ for $\delta_T,\,\delta_\mu$ and $n_{st.}=1$ for $\delta\rho_s$. Then, integrating along the radial variable, we find that, at leading order, 
\begin{align}
    \partial_{st.} \Phi_{[0]}(r)=\partial_{st.}\varrho_{h{[0]}} +2\,q_e^2 \int^r_0  dr \frac{e^{2g_c}}{U_c}\,a_c\,\delta\rho_{\ast(0)} \partial_{st.}\rho_{[0]}\,.
\end{align}
Using now \eqref{eq:mult_rho} once again, along with \eqref{eq:bulk_charge_der}, we arrive at the relations,
\begin{align}
    \frac{\partial_T \Phi_{[0]}(r)-\partial_T\varrho_{h[0]}}{\nu_{T\rho}^{[0]}}=\frac{\partial_\mu \Phi_{[0]}(r)-\partial_\mu\varrho_{h[0]}}{\nu_{\mu\rho}^{[0]}}=\frac{\partial_{\rho_s} \Phi_{[0]}(r)-\partial_{\rho_s}\varrho_{h[0]}}{\nu_{\rho\rho}^{[0]}}=\frac{\delta \Phi_{b (2)}(r)-\delta\varrho_{\ast h(2)}}{\delta\rho_{\ast (0)}^{v}}\,,
\end{align}
which hold everywhere in the bulk. Taking the near boundary limit $r\to\infty$ they give
\begin{align}\label{eq:susc_varrho_rel}
    \frac{\xi_{[0]}-\partial_T\varrho_{h[0]}}{\nu_{T\rho}^{[0]}}=\frac{\chi_{[0]}-\partial_\mu\varrho_{h[0]}}{\nu_{\mu\rho}^{[0]}}=\frac{\nu_{\mu\rho}^{[0]}-\partial_{\rho_s}\varrho_{h[0]}}{\nu_{\rho\rho}^{[0]}}=\frac{\Delta\varrho_{h(2)}}{\delta\rho_{\ast (0)}^{v}}\,.
\end{align}

\subsection{Expansions in the $s,\,\varrho,\,\rho_v$ ensemble}\label{Other_ens_scale}

The susceptibility relations above refer to the $T,\,\mu,\,\rho_s$ thermodynamic ensemble. In the $s,\,\varrho,\,\rho_v$ ensemble, the relations \eqref{eq:ch_ens} connecting the two ensembles, the $\ves$ expansions of Section \ref{sec:Background_linear} and the susceptibility relations \eqref{eq:susc_rho_0}, \eqref{eq:susc_varrho_rel} lead to the $\ves$ expansions,
\begin{align}\label{eq:susc_oe}
\partial_\varrho\phi^{(0)}&=\partial_\varrho\phi^{(0)}_{[0]}+\mathcal{O}(\ves^2)\,,\quad \partial_s\phi^{(0)}=\partial_s\phi^{(0)}_{[0]}+\mathcal{O}(\ves^2)\,,\quad \partial_{\rho_v}\phi^{(0)}=\ves \left(\partial_{\rho_v}\phi^{(0)}_{[0]}+\mathcal{O}(\ves^2)\right)\,,\nn
\partial_\varrho\rho^{(0)}&=\ves\left(\partial_\varrho\rho^{(0)}_{[2]}+\mathcal{O}(\ves^2)\right)\,,\quad \partial_s\rho^{(0)}=\ves\left(\partial_s\rho^{(0)}_{[2]}+\mathcal{O}(\ves^2)\right)\,,\quad \partial_{\rho_v}\rho^{(0)}=\frac{\delta \rho^{(0)}_{\ast(0)}}{\delta\rho^v_{\ast(0)}}+ \partial_{\rho_v}\rho^{(0)}_{[2]}\ves^2+\mathcal{O}(\ves^4)\,,\nn
    \partial_\varrho \varrho_h&=1+\partial_\varrho \varrho_{h[2]}\ves^2+\mathcal{O}(\ves^4)\,,\quad \partial_s\varrho_h=\ves^2 \partial_s\varrho_{h[2]}+\mathcal{O}(\ves^4)\,,\quad  \partial_{\rho_v} \varrho_h=-\frac{\Delta \varrho_{h(2)}}{\delta\rho^v_{\ast(0)}}+\partial_{\rho_v} \varrho_{h[2]}\ves^3+\mathcal{O}(\ves^5)\,.
\end{align}

\section{Further details on the symplectic current method}

In this appendix, we have gathered several technical details of the computation described in the main text in Section \ref{sec:Apply_SC}.

\subsection{Constraints from static perturbations}\label{sec:Sympl_constr}

First, we present certain first-order ODEs obeyed by the static perturbations of Section \ref{sec:Background_linear} and by the background solution at the critical point. These ODEs are used in the simplification of the bulk integrals $B_{{\delta_H},{\delta_{st.}}}$ introduced in equation \eqref{eq:Bulk_int_def}\footnote{Similar constraints were used in \cite{Donos:2022www} (see Appendix A therein). The only difference is that here we are interested in the ODEs obeyed by the leading parts of the static perturbations in the $\ves$ expansion. This means that we could find part of the constraints presented here by simply taking the near critical limit in the equations of Appendix A of \cite{Donos:2022www}.}.

The most efficient way to find these constraints is the symplectic current method. In particular, the symplectic current $P^\mu_{\delta_1,\delta_2}$ between any two static perturbations $\delta_1,\,\delta_2$ satisfies equation \eqref{eq:constr_deriv}. From $P^\mu_{\delta_{v_i},\delta_{m_i}},\,P^\mu_{\delta_{v_i},\delta_{s_{ti}}-\mu\,\delta _{m_i}},\,P^\mu_{\delta_T,\delta_{ s_{tt}}-\mu\,\delta_\mu},\,P^\mu_{\delta_\mu,\delta_{ s_{tt}}},\,P^\mu_{\ves\, \delta_{\rho_s},\delta_{ s_{tt}}-\mu\,\delta_\mu}$, $P^\mu_{\delta_T,\delta_{ s_{ii}}},\,P^\mu_{\delta_\mu,\delta_{ s_{ii}}},\,P^\mu_{\ves\,\delta_{\rho_s},\delta_{ s_{ii}}}$\footnote{Note that there is no summation implied over the spatial index $i$.}, at order $\ves^0$ we get, respectively,

\begin{align}\label{eq:stat_cons}
   & e^{2g_c}\,\tau_c\, a_c'=\varrho_c\,,\nn
   & e^{2g_c}\left(U_c'-2\,U_c\,g_c'-\tau_c\, a_c\, a_c'\right)=s_c\,T_c\,,\nn
   &e^{2g_c}\left(2\,g_c'\,\partial_T U_{[0]}-2\,\partial_T g_{[0]}\left(U_c'-2\,U_c\,g_c'\right)+U_c\left(4\,\partial_T g_{[0]}'+\delta\rho_{\ast(0)}'\partial_T \rho_{[0]}+\phi_c'\,\partial_T \phi_{[0]}\right)\right)=\nn&\quad=- a_c\,\partial_T \Phi_{[0]}-c_{\mu [0]}\,,\nn
   &e^{2g_c}\left(2\,g_c'\,\partial_\mu U_{[0]}-2\,\partial_\mu g_{[0]}\left(U_c'-2\,U_c\,g_c'\right)+U_c\left(4\,\partial_\mu g_{[0]}'+\delta\rho_{\ast(0)}'\partial_\mu \rho_{[0]}+\phi_c'\,\partial_\mu \phi_{[0]}\right)\right)=\nn&= -a_c\,\partial_\mu \Phi_{[0]}-T_c\,\xi_{[0]}\,,\nn
   & e^{2g_c}\left(2\,g_c'\partial_{\rho_s} U_{[0]}-2\,\partial_{\rho_s} g_{[0]}\left(U_c'-2U_c\,g_c'\right)+U_c\left(4\partial_{\rho_s} g_{[0]}'+\delta\rho_{\ast(0)}'\partial_{\rho_s} \rho_{[0]}+\phi_c'\,\partial_{\rho_s} \phi_{[0]}\right)\right)=\nn&=-a_c\,\partial_{\rho_s} \Phi_{[0]}-T_c\,\nu_{T\rho}^{[0]},\nn
   &e^{2g_c}\left(\tau_c\,a_c'\,\partial_T a_{[0]}-\partial_T U_{[0]}'-U_c\left(2\,\partial_T g_{[0]}'+\delta\rho_{\ast(0)}'\partial_{T} \rho_{[0]}+\phi_c'\,\partial_{T} \phi_{[0]}\right)\right)=-s_c\,,\nn
   &e^{2g_c}\left(\tau_c\,a_c'\,\partial_\mu a_{[0]}-\partial_\mu U_{[0]}'-U_c\left(2\,\partial_\mu g_{[0]}'+\delta\rho_{\ast(0)}'\partial_{\mu} \rho_{[0]}+\phi_c'\,\partial_{\mu} \phi_{[0]}\right)\right)=0\,,\nn
   &e^{2g_c}\left(\tau_c\,a_c'\,\partial_{\rho_s} a_{[0]}-\partial_{\rho_s} U_{[0]}'-U_c\left(2\,\partial_{\rho_s} g_{[0]}'+\delta\rho_{\ast(0)}'\partial_{\rho_s} \rho_{[0]}+\phi_c'\,\partial_{\rho_s} \phi_{[0]}\right)\right)=0\,.
\end{align}
Notice that the first of these constraints, evaluated on the horizon, gives
\begin{align}\label{eq:cons_tau}
    \frac{s_c}{4\pi}\tau_c^{(0)}a_c^{(0)}=\varrho_c\,.
\end{align}

Also, from $P^\mu_{\delta_{v_i},\delta_{m_i}}$ at order $\ves^2$ we get
\begin{align*}
   &\frac{\delta\Phi_{b(2)}}{2}-\delta f_{g[2]}(a_ca_c'\tau_c-U_c'+2 \,e^{2g_c}g_c')+e^{2g_c}\tau_ca_c'\delta f_{b[2]}-\delta f_{b[2]}'a_c U_c \tau_c-\delta f_{g[2]}'(U_c-e^{2g_c})\nn&=\frac{\delta\varrho_{\ast(2)}}{2}-\mu \chi_{JJ}^{[2]}\,.
\end{align*}

Finally, using the first equation of \eqref{eq:stat_cons} and evaluating the left-hand side of the above relation at the horizon we find
\begin{align}\label{eq:cons_fb2}
   \frac{s_c}{4\pi}\delta f_{g[2]}^{(1)}+\varrho_c\,\delta f_{b[2]}^{(0)}-\frac{\Delta\varrho_{h(2)}}{2} +\mu\, \chi_{JJ}^{[2]}=0\,.
\end{align}

\subsection{Divergences in $P^r_{\delta_H,\delta_{st.}}$}\label{app:diverg}

 Before integrating \eqref{eq:sympl_mom} from the horizon to infinity, we have to make sure that $P^r_{\delta_H,\delta_{st.}}$ is finite in the  $r\to\infty$ and $r\to 0$ limits.

Near the boundary, the following  symplectic current $r$-components are linearly diverging
\begin{align}
{P^r}^{[4]}_{\delta_H,\delta_\mu}\approx&-i\omega\chi_{[0]}\,\delta\theta_{[2](s)}(r+R)+\mathcal{O}\left((r+R)^0\right)\,,\nn
{P^r}^{[4]}_{\delta_H,\delta_T}\approx&-i\omega\xi_{[0]}\,\delta\theta_{[2](s)}(r+R)+\mathcal{O}\left((r+R)^0\right)\,,\\
{P^r}^{[4]}_{\delta_H,\ves\delta_{\rho_s}}\approx&-i\omega\nu_{\mu\rho}^{[0]}\,\delta\theta_{[2](s)}(r+R)+\mathcal{O}\left((r+R)^0\right)\,,\nn
{P^r}^{[4]}_{\delta_H,\delta_{v_i}}\approx&i k_i\,\varrho_c\,\delta\theta_{[2](s)}\,(r+R)+\mathcal{O}\left((r+R)^0\right)\,,\nn
{P^r}^{[4]}_{\delta_H,\delta_{s_{ij}}}\approx&-i\omega\varrho_c\,\delta\theta_{[2](s)}(r+R)\delta_{ij}+\mathcal{O}\left((r+R)^0\right)\,,\nn\notag
{P^r}^{[4]}_{\delta_H,\delta_{s_{it}}}\approx&-ik_i\varrho_c\,\delta\theta_{[2](s)}(r+R)+\mathcal{O}\left((r+R)^0\right)\,,
\end{align}
and all the rest are finite. If we denote as $P^{r[4]}_{\delta_H,\delta_{st.}}\approx c_{H,st.}(r+R)+\mathcal{O}\left((r+R)^0\right)$ the generic case, in order to take care of the diverging part, we manipulate \eqref{eq:sympl_mom} as
\begin{align}\label{eq:sympl_mom_mod}
   &\partial_r\left( {P^r}_{\delta_H,\delta_{st.}}-\ves^4\,c_{H,st.}\, r\right)-i\omega\ves^2 P^t_{\delta_H,\delta_{st.}}-i\omega\ves^2 S' P^r_{\delta_H,\delta_{st.}}+i\ves^2 k_i P^i_{\delta_H,\delta_{st.}}+\ves^4\,c_{H,st.}=0\Rightarrow\nn
   &\left( {P^r}_{\delta_H,\delta_{st.}}-\ves^4\,c_{H,st.}\, r\right)|^{r\to\infty}_{r\to 0}+B_{\delta_H,\delta_{st.}}=0\,,
\end{align}
with the bulk integral being
\begin{align}\label{eq:bulk_general}
B_{\delta_H,\delta_{st.}}=\int^\infty_0 dr\left( -i\omega\ves^2 P^t_{\delta_H,\delta_{st.}}-i\omega\ves^2 S' P^r_{\delta_H,\delta_{st.}}+i\ves^2 k_i P^i_{\delta_H,\delta_{st.}}+\ves^4\,c_{H,st.}\right)\,.
\end{align}

\subsection{Results from applying \eqref{eq:sympl_int} at order $\ves^4$}\label{sec:NLO_sympl}

In this subsection, we gather the results that one obtains by applying the symplectic current equation \eqref{eq:sympl_int} at order $\ves^4$. In several places we have made us of the thermodynamic relations of Section \ref{sec:setup} and equations \eqref{eq:cons_tau}, \eqref{eq:cons_fb2}. We have not written explicit formulas for the corresponding bulk integrals $B_{\delta_H,\delta_{st.}}$ because they are rather lengthy, but they can be obtained directly, by applying \eqref{eq:bulk_general}.

From the symplectic currents $P^\mu_{\delta_H,\delta_\mu}$, $P^\mu_{\delta_H,\delta_{m_i}}$ and $P^\mu_{\delta_H,\ves\delta_{\rho_s}}$ we find, respectively, \begin{align}\label{eq:Jt_nlo}&
    \delta J^t_{[4]}\ves^4=\chi_{[0]} \left(\delta m_{t[4]}\ves^4+\frac{\delta \mu_{\ast(2)}}{2}\ves^2 \delta s_{tt}\right)+\chi_{[2]}\ves^2\left(\delta m_{t[2]}\ves^2+\mu\delta s_{tt}\right)+\xi_{[2]}\ves^2\delta T\nn&+\frac{\delta\varrho_{\ast(2)}}{2}\ves^2\delta s^{tt}+\nu_{\mu\rho}^{[2]}\ves\,\delta\rho_{[3](s)}\ves^3-i\omega\delta\theta_{[2](s)}\ves^2\chi_{[0]} R\nn&-\frac{i \omega\ves^2\, \xi_{[0]}}{4\pi\, s_c}\left(\frac{c_{\mu[0]}}{T_c}\delta T+\xi_{[0]}\,(\delta\mu+\mu\, \delta s_{tt})+\frac{\nu_{T\rho}^{[0]}}{\ves}\,\delta\pi+s_c\,\delta s_{ij}\delta^{ij}-s_c\,\delta s_{tt}\right)\nn&+i\omega\ves^2\frac{s_c}{4\pi}\partial_\mu\phi^{(0)}_{[0]}\left( \partial_T\phi^{(0)}_{[0]}\,\delta T+ \partial_\mu\phi^{(0)}_{[0]}\,\left(\delta \mu+\mu\delta s_{tt}\right)+\frac{\partial_{\rho_s}\phi^{(0)}_{[0]}}{\ves}\,\delta \pi\right)\nn&+\frac{i\omega\ves^2}{4\pi}\left(s_c\,\frac{\partial_\mu\rho^{(0)}_{[0]}}{\ves} \delta\rho^{(0)}_{[3]}\ves^3+s_c\,\partial_\mu\rho_{[2]}^{(0)}\ves\,\delta\rho_{[1]}^{(0)}\ves+\frac{\delta s_{\ast(2)}}{2}\ves^2\frac{\partial_\mu\rho_{[0]}^{(0)}}{\ves}\delta\rho_{[1]}^{(0)}\ves\right)\nn&
-\frac{T_c\,\xi_{[0]}}{2}\left(\delta g_{tt[4]}^{(0)}+\delta g_{rr[4]}^{(0)}\right)\ves^4+\xi_{[0]}\frac{\delta T_{\ast(2)}}{2}\ves^2\frac{\delta T}{T_c}-\partial_\mu \varrho_{h[0]}\delta B_{t[4]}^{(0)}\ves^4-\partial_\mu \varrho_{h[2]}\ves^2\delta\mu_r+B_{\delta_H,\delta_{\mu}}^{[4]}\ves^4\,,
\end{align}

\begin{align}\label{eq:Ji_nlo}
    &\delta J^i_{[4]}\ves^4=- \chi_{JJ}^{[2]}\ves^2\delta  m^i_{[2]}\ves^2+\frac{\delta\varrho_{\ast(2)}}{2}\ves^2\delta s^{it}+\left(\frac{\delta\varrho_{\ast(2)}}{2}-\mu\chi_{JJ}^{[2]}\right)\ves^2\delta v^i\nn&+\tau^{(0)}_c\left(ik_i\ves^2\delta\mu_r+i\omega\ves^2 (\delta m^i_{[2]}\ves^2+\mu\delta v^i)\right)-\tau_c^{(0)}a^{(0)}_c \left(\delta g_{ti[4]}^{(0)}\ves^4+\frac{\delta s_{\ast(2)}}{8\pi}\ves^2\left(\delta v^i+\delta s^{it}\right)\right)+B^{[4]}_{\delta_H,\delta m_i}\ves^4\,.
    \end{align}

\begin{align}\label{eq:ampl_nlo}&
 -\delta\rho_{[3](v)}\ves^3+\frac{\nu_{\mu\rho}^{[0]}}{\ves} \left(\delta m_{t[4]}\ves^4+\frac{\delta \mu_{\ast(2)}}{2}\ves^2\delta s_{tt}\right)+\nu_{\mu\rho}^{[2]}\ves\left(\delta m_{t[2]}\ves^2+\mu\delta s_{tt}\right)+\nu_{T\rho}^{[2]}\ves\delta T+\nu_{\rho\rho}^{[2]}\delta\rho_{[3](s)}\ves^3\nn&-i\omega\ves^2\delta\theta_{[2](s)}\ves^2\frac{\nu_{\mu\rho}^{[0]}}{\ves} R-\frac{i \omega\ves^2}{4\pi\, s_c}\frac{\nu_{T\rho}^{[0]}}{\ves}\left(\frac{c_{\mu[0]}}{T_c}\delta T+\xi_{[0]}\,(\delta\mu+\mu\, \delta s_{tt})+\frac{\nu_{T\rho}^{[0]}}{\ves}\,\delta\pi+s_c\,\delta s_{ij}\delta^{ij}-s_c\,\delta s_{tt}\right)\nn&+i\omega\ves^2\frac{s_c}{4\pi}\frac{\partial_{\rho_s}\phi^{(0)}_{[0]}}{\ves}\left(\partial_T\phi^{(0)}_{[0]}\delta T+\partial_\mu\phi^{(0)}_{[0]}\left(\delta \mu+\mu \delta s_{tt}\right)+\frac{\partial_{\rho_s}\phi^{(0)}_{[0]}}{\ves}\delta\pi\right)\nn&+\frac{i\omega\ves^2}{4\pi}\left(s_c\,\frac{\partial_{\rho_s}\rho^{(0)}_{[0]}}{\ves^2} \delta\rho_{[3]}^{(0)}\ves^3+s_c\,\partial_{\rho_s}\rho_{[2]}^{(0)}\delta\rho_{[1]}^{(0)}\ves+\frac{\delta s_{\ast(2)}}{2}\ves^2\frac{\partial_{\rho_s}\rho_{[0]}^{(0)}}{\ves^2}\delta\rho_{[1]}^{(0)}\ves\right)\nn& -\frac{T_c\,\frac{\nu_{T\rho}^{[0]}}{\ves}}{2}\left(\delta g_{tt[4]}^{(0)}+\delta g_{rr[4]}^{(0)}\right)\ves^4+\frac{\nu_{T\rho}^{[0]}}{\ves}\frac{\delta T_{\ast(2)}}{2}\ves^2\frac{\delta T}{T_c}-\frac{\partial_{\rho_s} \varrho_{h[0]}}{\ves}\delta B_{t[4]}^{(0)}\ves^4-\partial_{\rho_s} \varrho_{h[2]}\ves\delta\mu_r+B_{\delta_H,\delta_{\rho_s}}^{[3]}\ves^3=0\,.
\end{align}

From $P^\mu_{\delta_H,\delta_T}$ and $P^\mu_{\delta_H,\delta_{v_i}}$ we get, respectively, the following constraints,
\begin{align}\label{eq:temp_constr}
 &\xi_{[0]} \left(\delta m_{t[4]}\ves^4+\frac{\delta \mu_{\ast(2)}}{2}\ves^2\delta s_{tt}\right)+\xi_{[2]}\ves^2\left(\delta m_{t[2]}\ves^2+\mu \delta s_{tt}\right)+\frac{c_{\mu[2]}\ves^2}{T_c}\delta T+\nu_{T\rho}^{[2]}\ves\,\delta\rho_{[3](s)}\ves^3\nn&-i\omega\ves^2\delta\theta_{[2](s)}\ves^2\,\xi_{[0]} R-\frac{i \omega\ves^2}{4\pi\, s_c}\frac{c_{\mu[0]}}{T_c}\left(\frac{c_{\mu[0]}}{T_c}\delta T+\xi_{[0]}\,(\delta\mu+\mu\, \delta s_{tt})+\frac{\nu_{T\rho}^{[0]}}{\ves}\,\delta\pi+s_c\,\delta s_{ij}\delta^{ij}-s_c\,\delta s_{tt}\right)\nn&+i\omega\ves^2\frac{s_c}{4\pi}\partial_{T}\phi^{(0)}_{[0]}\left(\partial_{T}\phi^{(0)}_{[0]} \delta T+\partial_{\mu}\phi^{(0)}_{[0]} (\delta \mu+\mu \delta s_{tt})+\frac{\partial_{\rho_s}\phi^{(0)}_{[0]}}{\ves}\delta \pi\right)\nn&+\frac{i\omega\ves^2}{4\pi}\left(s_c\,\frac{\partial_T\rho^{(0)}_{[0]} }{\ves}\delta\rho_{[3]}^{(0)}\ves^3+s_c\,\partial_T\rho_{[2]}^{(0)}\ves\delta\rho_{[1]}^{(0)}\ves+\frac{\delta s_{\ast(2)}}{2}\ves^2\frac{\partial_T\rho_{[0]}^{(0)}}{\ves}\delta\rho_{[1]}^{(0)}\ves\right)\nn& -\frac{c_{\mu[0]}}{2}\left(\delta g_{tt[4]}^{(0)}+\delta g_{rr[4]}^{(0)}\right)\ves^4-2\pi\delta^{ij}\delta g_{ij [4]}^{(0)}\ves^4+\frac{\delta s_{\ast(2)}}{2}\ves^2\delta^{ij}\delta s_{ij}\nn&-\partial_T \varrho_{h[0]}\delta B_{t[4]}^{(0)}\ves^4-\partial_T \varrho_{h[2]}\ves^2\delta\mu_r+B_{\delta_H,\delta_T}^{[4]}\ves^4=0\,,
\end{align}

\begin{align}\label{eq:veloc_constr}
& \varrho_c\, \delta m_{i[4]}\ves^4+\left(\frac{\Delta\varrho_{h(2)}}{2}\ves^2-\mu\chi_{JJ}^{[2]}\ves^2\right)\left(\delta m_{i[2]}\ves^2+\mu\, \delta v_i\right)+\left(s_c\frac{\delta T_{\ast(2)}}{2}\ves^2+\varrho_c \frac{\delta \mu_{\ast(2)}}{2}\ves^2\right)\delta v_i\nn
    &+i k_i\ves^2\varrho_c \delta\theta_{[2](s)}\ves^2R-\tau^{(0)}_c\left(a_c^{(0)}\frac{s_c}{4\pi}\delta B_{i[4]}^{(0)}\ves^4+a_c^{(0)}\mu\,\delta g_{ti[4]}^{(0)}\ves^4-i \omega\ves^2 \mu \left(\delta m_{i[2]}\ves^2+\mu\,\delta v_i\right)-i\mu\, k_i\ves^2\delta\mu_r\right)\nn&
    +\frac{ik_i\ves^2}{4\pi}\left(\frac{c_{\mu[0]}}{T_c}\delta T+\xi_{[0]}\left(\delta\mu+\mu\, \delta s_{tt}\right)+\frac{\nu_{T\rho}^{[0]}}{\ves}\,\delta\pi\right) -i\omega\ves^2\frac{s_c}{4\pi}\delta f_{c [0]}^{(0)}\left(\delta m_{i[2]}\ves^2+\mu\,\delta v_i\right)\nn&
   +\frac{s_c}{4\pi}\left(i\omega\ves^2\delta v_i+i k_i\ves^2\, \delta s_{jk}\delta^{jk}+2\,g_c^{(1)}\,\delta g_{ti[4]}^{(0)}\ves^4-\delta g_{ti[4]}^{(1)}\ves^4\right)-4\pi T_c \delta g_{ti[4]}^{(0)}\ves^4\nn&
   +\frac{\delta s_{\ast(2)}}{2}\ves^2\frac{s_c T_c+\mu \varrho_c}{s_c}\left(\delta s_{it}-\delta v_i\right)+\frac{s_c^2}{16\pi^2}\delta g_{\ast(2)}^{(1)}\ves^2\left(\delta s_{it}-\delta v_i\right)+B^{[4]}_{\delta_H,\delta_{v_i}}\ves^4=0\,.
\end{align}

 From $P^\mu_{\delta_H,\delta_{s_{tt}}-\mu\delta\mu},\,P^\mu_{\delta_H,\delta_{s_{ti}}},\,P^\mu_{\delta_H,\delta_{s_{ij}}}$ and $P^\mu_{\delta_H,\delta_{s_{it}}}$ we obtain
\begin{align}\label{eq:Ttt_nlo}
    &\delta T^{tt}_{[4]}\ves^4=\mu\, \delta J^t_{[4]}\ves^4+\xi_{[0]}\frac{\delta \mu_{\ast(2)}}{2}\ves^2\delta T+\left(\frac{\delta\mu_{\ast(2)}}{2}\ves^2\chi_{[0]}+\frac{\delta T_{\ast(2)}}{2}\ves^2\xi_{[0]}\right)\left(\delta\mu+\mu\delta s_{tt}\right)+\frac{\delta \epsilon_{\ast(2)}}{2}\ves^2\,2\delta s^{tt}\nn&-\delta\rho_{\ast(0)}^v \ves\,\delta\rho_{[3](s)}\ves^3+\left(\frac{\nu_{\mu\rho}^{[0]}}{\ves}\frac{\delta \mu_{\ast(2)}}{2}\ves^2+\frac{\nu_{T\rho}^{[0]}}{\ves}\frac{\delta T_{\ast(2)}}{2}\ves^2\right)\delta\pi\nn&+i\frac{s_c}{4\pi}\left(\omega\ves^2\delta s_{i j}\delta^{ij}-k_i \ves^2\left(\delta s^{i t}+\delta v^i\right)\right)+\frac{i\omega\ves^2}{4\pi}\left(\frac{c_\mu^{[0]}}{T_c}\delta T+\xi_{[0]}\left(\delta \mu+\mu \delta s_{tt}\right)+\frac{\nu_{T\rho}^{[0]}}{\ves}\delta\pi\right)\nn&+\frac{c_\mu^{[0]}}{T_c}\frac{\delta T_{\ast(2)}}{2}\ves^2\delta T-\mu \frac{\delta\varrho_{\ast(2)}}{2}\ves^2\delta s^{tt}
    -T_c\frac{\delta s_{\ast(2)}}{2}\ves^2\delta^{ij}\delta s_{ij}+2\pi T_c \delta g_{ij[4]}^{(0)}\delta^{ij}\ves^4+B_{\delta_H,\delta_{s_{tt}}-\mu\delta_\mu}^{[4]}\ves^4\,,
\end{align}
\begin{align}\label{eq:Tti_nlo}
    &\delta T^{ti}_{[4]}\ves^4=-\mu\chi_{JJ}^{[2]}\ves^2\delta m^i_{[2]}\ves^2+\left(\frac{\delta\epsilon_{\ast(2)}}{2}+\frac{\delta p_{\ast(2)}}{2}-\mu^2 \chi_{JJ}^{[2]}\right)\ves^2\delta v^i+\frac{\delta\epsilon_{\ast(2)}}{2}\ves^2\delta s^{it}\nn&+i\mu\tau^{(0)}_c\left(\omega\ves^2\left(\delta m^i_{[2]}\ves^2+\mu\delta v^i\right)+k^i\ves^2\delta\mu_r\right)\nn&-\frac{4\pi}{s_c}\left(s_c T_c+\mu\varrho_c\right)\,\left(\delta g_{ti[4]}^{(0)}\ves^4+\frac{\delta s_{\ast(2)}}{8\pi}\ves^2\left(\delta v^i+\delta s^{it}\right)\right)+B^{[4]}_{\delta_H,\delta_{s_{ti}}}\ves^4\,,
\end{align}

\begin{align}\label{eq:Tij_nlo}
    \delta &T^{ij}_{[4]}\ves^4=-2\frac{\delta p_{\ast(2)}}{2}\ves^2\delta s^{(ij)}+\left(\frac{\delta s_{\ast(2)}}{2}\delta T+\frac{\delta \varrho_{\ast(2)}}{2}\left(\delta\mu+\mu\delta s_{tt}\right)+\varrho_c \frac{\delta \mu_{\ast(2)}}{2}\delta s_{tt}\right)\ves^2\delta^{ij}\nn&+\delta\rho_{\ast(0)}^v\ves\delta\rho_{[3](s)}\ves^3\delta^{ij}-\frac{i\omega\ves^2}{4\pi}\left(\frac{c_{\mu[0]}}{T_c}\delta T+\xi_{[0]} \left(\delta \mu+\mu \delta s_{tt}\right)+\frac{\nu_{T\rho}^{[0]}}{\ves}\,\delta\pi\right)\delta^{ij}\nn&
    +\frac{i s_c }{4\pi}\left(2  k_{(i}\left(\delta s_{j)t}-\delta v_{j)}\right)-k_l\left(\delta s_{lt}-\delta v_l\right)\delta_{ij}+\omega \delta s_{tt}\delta_{ij}+2\omega \delta s_{(ij)}-2\omega\delta s_{kl}\delta^{kl}\delta_{ij}\right)\ves^2\nn
    &+\varrho_c \left(\delta m_{t[4]}\ves^4-i\omega\ves^2\delta\theta_{[2](s)} \ves^2R\right)\delta^{ij}+\frac{\Delta\varrho_{h(2)}}{2}\ves^2\delta\mu_r\delta^{ij}\nn&-\frac{s_c T_c}{2}\left(\delta g_{tt[4]}^{(0)}+\delta g_{rr[4]}^{(0)}\right)\ves^4\delta^{ij}+s_c\frac{\delta T_{\ast(2)}}{2}\ves^2\frac{\delta T}{T_c}\delta^{ij}
    -\frac{s_c}{4\pi} a_c^{(0)}\tau^{(0)}_c\delta B_{t[4]}^{(0)}\ves^4\delta^{ij}+B^{[4]}_{\delta_H,\delta_{s_{ij}}}\ves^4\,,
\end{align}

\begin{align}\label{eq:Tti_nlo_v2}
    &\delta T^{ti}_{[4]}\ves^4=\frac{\delta\epsilon_{\ast(2)}}{2}\ves^2\delta s^{it}+\left(T_c\frac{\delta s_{\ast(2)}}{2}\ves^2+\mu \frac{\delta \varrho_{\ast(2)}}{2}\ves^2\right)\delta v^i-i\frac{s_c}{4\pi}\left(\omega\ves^2\,\delta v^i+k^i\ves^2 \delta^{jk}\delta s_{jk}\right)\nn&+i\omega\ves^2\delta f_{c[0]}^{(0)} \frac{s_c}{4\pi
    }\left(\delta m^i_{[2]}\ves^2+\mu\, \delta v^i\right)-\varrho_c\,\delta m^i_{[4]}\ves^4-ik^i\ves^2\varrho_c\,\delta\theta_{[2](s)}\ves^2R\nn&-\frac{\Delta\varrho_{h(2)}}{2}\ves^2\left(\delta m^i_{[2]}\ves^2+\mu \delta v^i\right)-\frac{ik^i\ves^2}{4\pi}\left(\frac{c_{\mu[0]}}{T_c}\delta T+\xi_{[0]}\left(\delta\mu+\mu\delta s_{tt}\right)+\frac{\nu_{T\rho}^{[0]}}{\ves}\delta\pi\,\right)\nn&+\frac{s_c}{4\pi}a_c^{(0)}\tau_c^{(0)}\delta B_{i[4]}^{(0)}\ves^4+\frac{s_c}{4\pi}\left(\delta g_{ti[4]}^{(1)}-2g_c^{(1)}\delta g_{ti[4]}^{(0)}\right)\ves^4+\frac{s_c^2}{16\pi^2}\left(\delta s^{it}+\delta v^i\right)\delta g_{\ast(2)}^{(1)}\ves^2+B^{[4]}_{\delta_H,\delta_{s_{it}}}\ves^4\,.
\end{align}

\subsection{Determination of $\delta\rho_{[3]}^{(0)}$}\label{app:delta_rho_0}

In this appendix, we describe how to find an expression for the horizon constant $\delta\rho_{[3]}^{(0)}$.
We first need to consider a spacetime-independent perturbation $\hat{\delta} f(r)$ defined as
\begin{align}
    \hat{\delta} f=\delta_T f\, \delta T+\delta_\mu f\, \delta \mu+\delta_{\rho_s} f\, \delta \pi+\delta_{s_{tt}}f\, \delta s_{tt}+\delta_{s_{xx}}f\, \delta s_{xx}+\delta_{s_{yy}}f\, \delta s_{yy}\,.
\end{align}
It is important that the parameters $\delta T, \delta\mu,\, \delta\pi,\,\delta s_{tt},\,\delta s_{xx},\,\delta s_{yy}$ are precisely those of the hydrodynamic perturbation \eqref{eq:hydro_exp}. Note also that this static perturbation has a source for the amplitude at order $\ves^3$, chosen to be equal to $\delta\pi$, given by \eqref{eq:rho_vev_lead}.

Then we have to construct the symplectic currents $P^\mu_{\delta_H,\,\ves\delta_{\rho_s}}$ and $P^\mu_{\hat{\delta},\,\ves\delta_{\rho_s}}$. Applying \eqref{eq:constr_deriv} and \eqref{eq:sympl_mom}  leads to 
\begin{align}\label{eq:pr_rhos}
P^{r[2]}_{\hat{\delta},\,\ves\delta_{\rho_s}}(r)&=\lim_{r\to 0}P^{r[2]}_{\hat{\delta},\,\ves\delta_{\rho_s}}(r)\,,\nn
    P^{r[2]}_{\delta_H,\,\ves\delta_{\rho_s}}(r)&=\lim_{r\to 0}P^{r[2]}_{\delta_H,\,\ves\delta_{\rho_s}}(r)\,.
\end{align}
Subtracting these two equations and substituting the explicit formulas for the $r$-components of the symplectic currents, yields  the following ODE for $\delta \rho_{[3]}$,
\begin{align}\label{eq:rhos_ode}
     \delta\rho_{\ast (0)}\ves\,(-\delta\rho'_{[3]}\,\ves^3+\hat{\delta}\rho'_{[3]}\,\ves^3)+  \delta\rho'_{\ast (0)}\ves\,(\delta\rho_{[3]}\,\ves^3-\hat{\delta}\rho_{[3]}\,\ves^3)+X_\rho=0\,,
\end{align}
where we have introduced the bulk function
\begin{align}\label{eq:X_expr}
    X_\rho=\frac{1}{e^{2g_c} U_c}\left( \delta\mu_r(\delta\varrho_{h\ast(2)}\ves^2-\delta \Phi_{b(2)}\ves^2)+i \omega\,\ves^2  \frac{\delta\rho_{[1](v)}}{\delta\rho_{\ast(0)}^v} \left(e^{2g_c}\,U_c\, S'  \,\delta\rho_{\ast (0)}^2\ves^2-\frac{s_c}{4 \pi}  \left(\delta\rho^{(0)}_{\ast (0)}\right)^2\ves^2\right)\right)\,.
\end{align}
In writing \eqref{eq:rhos_ode} and \eqref{eq:X_expr} we have employed \eqref{eq:ampl_mult} and \eqref{eq:mult_rho}. Integrating \eqref{eq:rhos_ode} from the horizon to a point in the interior with radial coordinate $r$ we get
\begin{equation}\label{eq:rho_ode_v2}
   \frac{ \delta\rho_{[3]}(r)\,\ves^3-\hat{\delta}\rho_{[3]}(r)\,\ves^3}{\delta\rho_{\ast (0)}\,\ves}-\frac{ \delta\rho^{(0)}_{[3]}\,\ves^3-\hat{\delta}\rho^{(0)}_{[3]}\,\ves^3}{\delta\rho_{\ast (0)}^{(0)}\ves}=\int^r_0 dr' \frac{X_\rho}{\delta\rho_{\ast (0)}^2\,\ves^2}\,.
\end{equation}

Now we would like to take the near boundary limit of this equation. Using \eqref{eq:ampl_lead}, we can show that the integrand on the right-hand side behaves as
\begin{equation}
    \frac{X_\rho}{\delta\rho_{\ast (0)}^2\,\ves^2}=\frac{\delta\rho_{[3](s)}\,\ves^3-\delta\pi}{\delta\rho_{\ast (0)}^v\,\ves}+\mathcal{O}\left(\frac{1}{(r+R)^2}\right)\,,
\end{equation}
for $r\to \infty$. Then from \eqref{eq:rho_ode_v2}
\begin{align}
    \frac{ \delta\rho_{[3]}\,\ves^3-\hat{\delta}\rho_{[3]}\,\ves^3}{\delta\rho_{\ast (0)}\,\ves}-\frac{ \delta\rho^{(0)}_{[3]}\,\ves^3-\hat{\delta}\rho^{(0)}_{[3]}\,\ves^3}{\delta\rho_{\ast (0)}^{(0)}\,\ves}+r\frac{\delta\pi-\delta\rho_{[3](s)}\,\ves^3}{\delta\rho_{\ast(0)}^v\,\ves}=\int^r_0 dr' \left(\frac{X_\rho}{\delta\rho_{\ast (0)}^2\,\ves^2}+\frac{\delta\pi-\delta\rho_{[3](s)}\,\ves^3}{\delta\rho_{\ast(0)}^v\,\ves}\right)\,
\end{align}
and taking the  $r \to \infty$ limit
\begin{align}\label{eq:rho_0_fin}
\delta\rho_{[3](v)}\,\ves^3&=\hat{\delta}\rho_{[3]v}\,\ves^3+ \frac{\delta\rho_{\ast (0)}^v}{\delta\rho_{\ast (0)}^{(0)}}\left( \delta\rho^{(0)}_{[3]}\,\ves^3-\hat{\delta}\rho_{[3]}^{(0)}\,\ves^3\right)-R\left(\delta\rho_{[3](s)}\,\ves^3-\delta\pi\right)\nn
&+\delta\rho_{\ast (0)}^v\,\ves\int^\infty_0 dr'\left(\frac{X_\rho}{\delta\rho_{\ast (0)}^2\,\ves^2}+\frac{\delta\pi-\delta\rho_{[3](s)}\,\ves^3}{\delta\rho_{\ast(0)}^v\,\ves}\right)\,.
\end{align}

This last equation, allows us to express $\delta\rho_{[3]}^{(0)}$ in terms of $\delta\rho_{[3](v)}$ and background quantities. In particular, we note that
\begin{align}
\hat{\delta}\rho_{[3]v}\,\ves^3&=\nu_{T\rho}^{[2]}\,\ves\,\delta T+\nu_{\mu\rho}^{[2]}\,\ves\left(\delta\mu+\mu\,\delta s_{tt}\right)+\nu_{\rho\rho}^{[2]}\,\delta \pi\,,\nn
    \hat{\delta} \rho_{[3]}^{(0)}\,\ves^3&=\partial_T \rho^{(0)}_{[2]}\,\ves\,\delta T+\partial_\mu \rho^{(0)}_{[2]}\,\ves\left(\delta \mu+\mu\,\delta s_{tt}\right)+\partial_{\rho_s}\rho^{(0)}_{[2]}\,\delta\pi\,.
\end{align}

Closing this subsection, let us just comment that another way to find $\delta\rho_{[3]}^{(0)}$ would be to start with the second equation of \eqref{eq:pr_rhos} and simplify the left-hand side using constraints similar to those in \ref{sec:Sympl_constr}, obtained by considering $P^\mu_{\delta_\mu,\,\ves \delta\rho_s}$, $P^\mu_{\delta_T,\,\ves \delta\rho_s}$.

\subsection{Determination of $\delta\mu_r$ and $\delta B_{t[4]}^{(0)}$}\label{app:delta_B_0}

In previous work (see, e.g., \cite{Donos:2022qao}), in order to determine the $t$-component of the gauge field at the horizon, $\delta B_t^{(0)}$, we simply took the near horizon limit of  the $r$-component gauge field equation and used the third regularity condition \eqref{eq:nh_reg}. In our case, this condition implies (see \eqref{eq:hor_reg_eps})
\begin{align}\label{eq:reg_hor_hydro}
    \delta B_{r[2]}\ves^2&=\frac{\delta\mu_r}{4\pi T_c\,r}+\mathcal{O}(r^0)\,,\nn
    \delta B_{r[4]}\ves^4&=\frac{\delta B_{t[4]}^{(0)}\,\ves^4}{4\pi T_c\,r}-\frac{\delta T_{\ast(2)}\,\ves^2\delta\mu_r}{8\pi T_c^2\,r}+\mathcal{O}(r^0)\,.
\end{align}

We could, indeed, apply this method to find $\delta\mu_r$ here. However, this method cannot give us $\delta B_{t[4]}^{(0)}$ since the resulting equation involves $\delta B_{t[4]}^{(1)}$ as well, which is also undetermined. Moreover, we highlight that several bulk integrals $B_{\delta_H,\delta_{st.}}$ contain $\delta B_{r[2]}(r)$ in their integrands, so we need to find an explicit expression for it as well.

In order to find all of $\delta B_{r[2]}(r),\,\delta\mu_r,\,\,\delta B_{t[4]}^{(0)}$ we can use the last equation of motion in \eqref{eq:eom}\footnote{As a general remark, note that this equation is redundant, since it follows from the gauge field equation by taking a total divergence.}. More specifically, this equation can be viewed as a first-order ODE for $\delta B_r$ which can be easily solved. The resulting constant of integration is fixed by imposing the fall-off \eqref{eq:pert_uv_bcs} for $\delta B_r$ at $r\to\infty$. Then, taking the near horizon limit of the solution for $\delta B_r$ and applying \eqref{eq:reg_hor_hydro} we can read $\delta B_t^{(0)}$ at each order in $\ves$.

Considering the last of \eqref{eq:eom} at order $\ves^4$, we find 
\begin{align}\label{eq:Br_2}
    \delta B_{r[2]}(r)\ves^2&=-i\omega\ves^2\, \frac{\delta\rho_{[1](v)}}{\delta\rho_{\ast(0)}^v} \frac{\delta \Phi_{b(2)}-\delta\varrho_{\ast(2)}}{q_e^2\, e^{2g_c}\,U_c\, \delta\rho_{\ast(0)}^2}-\frac{a_c\,S'}{U_c}\left(\delta s_{tt}\,U_c-\delta  T\,\partial_T\,U_{[0]}\right)\nn&\quad+ \frac{\left(\delta\rho_{\ast(0)}^v\right)^2}{ e^{2g_c}\,U_c\, \delta\rho_{\ast(0)}^2}\delta\theta_{[2](s)}\ves^2\,,
\end{align}
and imposing \eqref{eq:reg_hor_hydro} as $r\to 0$, that
\begin{align}
    \delta\mu_r=\frac{4\pi}{q_e^2\, s_c\left(\delta\rho_{\ast(0)}^{(0)}\right)^2}\left(i\omega\ves^2\frac{\delta\rho_{[1](v)}}{\delta\rho_{\ast(0)}^v}\Delta\varrho_{h(2)}+q_e^2\left(\delta\rho_{\ast(0)}^v\right)^2\delta\theta_{[2](s)}\ves^2\right)\,.
\end{align}

Similarly, at the next order in $\ves$, using the results of  Appendix \ref{app:delta_rho_0} and equations \eqref{eq:bulk_charge_der}, \eqref{eq:chi_JJ_rel}, \eqref{eq:Br_2}, we find

\begin{align}
   & \delta B_{t[4]}^{(0)}\,\ves^4=\frac{4\pi}{q_e^2 s_c \left(\delta\rho_{\ast(0)}^{(0)}\ves\right)^2}
    \left(i\,k_i\,\ves^2\,\chi_{JJ}^{[2]}\,\ves^2\left(\delta m^i+\mu\, \delta v^i\right)+ \frac{\Delta\varrho_{h(2)}}{2 }\ves^2\left(i\,\omega\ves^2\, \delta^{ij}\delta s_{ij}-i\,k_i\,\ves^2\left(\delta s^{it}+\delta v^i\right)\right)\right)\nn
    &+\frac{1}{s_c \left(\delta\rho_{\ast(0)}^{(0)}\ves\right)^2}\left(8\pi\,\delta\theta_{[2](s)}\ves^2\delta\rho_{\ast(0)}^v\ves \frac{\delta\rho_{\ast(2)}^v}{3!}\ves^3- \delta\mu_r\, \delta\rho_{\ast(0)}^{(0)}\ves\left(\frac{\delta s_{\ast(2)}}{2}\ves^2\delta\rho_{\ast(0)}^{(0)}\ves+2s_c \frac{\delta\rho_{\ast(2)}^{(0)}}{3!}\ves^3\right)\right)+\nn
    &\frac{4\pi\,i\omega\ves^2  }{q_e^2\,s_c \left(\delta\rho_{\ast(0)}^{(0)}\ves\right)^2}\left(\left(\xi_{[2]}-\partial_T\varrho_{h[2]}-\Delta\varrho_{h(2)}\frac{\nu_{T\rho}^{[2]}}{\delta\rho_{\ast(0)}^v}\right)\ves^2\delta T+\left(\chi_{[2]}-\partial_\mu\varrho_{h[2]}-\Delta\varrho_{h(2)}\frac{\nu_{\mu\rho}^{[2]}}{\delta\rho_{\ast(0)}^v}\right)\ves^2\left(\delta\mu+\mu \delta s_{tt}\right)\right.\nn&\left.+\left(\nu_{\mu\rho}^{[2]}-\partial_{\rho_s}\varrho_{h[2]}-\Delta\varrho_{h(2)}\frac{\nu_{\rho\rho}^{[2]}}{\delta\rho_{\ast(0)}^v}\right)\ves\,\delta \pi+\frac{\Delta\varrho_{h(2)}}{\delta\rho_{\ast(0)}^v}\ves\left(\delta\rho_{[3](v)}\ves^3+R\left(\delta\rho_{[3](s)}\ves^3-\delta\pi\right)\right)+BI_0\right)\,,
\end{align}
where $BI_0$ is the bulk integral
\begin{align}
BI_0&=\int^\infty_0dr'\left(\left(\delta\varrho_{\ast h(2)}-\delta\Phi_{b(2)}\right)\frac{X_\rho}{\delta\rho_{\ast(0)}^2}+\frac{\Delta\varrho_{h(2)}}{\delta\rho_{\ast(0)}^v}\ves\left(\delta\rho_{[3](s)}\ves^3-\delta\pi\right)+q_e^2 \frac{e^{2g_c}}{U_c}\left(\delta\rho_{\ast(0)}\ves\right)^2\delta\mu_r\right.\nn&\left. -q_e^2 \delta\theta_{[2](s)}\ves^2\left(\delta\rho_{\ast(0)}^v\ves\right)^2S'+i\omega\ves^2 \left(\delta \Phi_{b(2)}-\delta \varrho_{\ast(2)}\right)\ves^2\frac{\delta\rho_{[1](v)}}{\delta\rho_{\ast(0)}^v}S'\right)\,.
\end{align}

\section{Origin of the $Z_\pi$ coefficient}\label{app:Zpi}

In this appendix, we will seek the origin of the complex coefficient $Z_\pi$, which enters the equation of motion for the order parameter \eqref{eq:ord_linear}. In addition, we present an alternative method to find explicit expressions \eqref{eq:ReZ_pi}\eqref{eq:ImZ_pi}, independent of the symplectic current.  

We will consider our holographic system exactly at the critical point, where the complex field $\psi$ is trivial in the background, and its linear fluctuations decouple from the rest of the bulk fields. The bulk equations of motion for the real and imaginary parts of $\delta\psi=\delta\psi_R+i\, \delta\psi_I$ read as follows,
\begin{align}\label{eq:Critic_eqs}
    \partial_\mu \left( e^{2 g_c} g^{\mu\nu}_c\partial_\nu \delta\psi_R\right)+2 q_e e^{2g_c}\frac{a_c}{U_c}\partial_t \delta\psi_I+e^{2g_c}\left(q_e^2 \frac{a_c^2}{U_c}-2\partial_{|\psi|^2 }V_c-\partial_{|\psi|^2 }\tau_c F^2_c\right)\delta\psi_R=0\,,\nn
    \partial_\mu \left( e^{2 g_c} g^{\mu\nu}_c\partial_\nu \delta\psi_I\right)-2 q_e e^{2g_c}\frac{a_c}{U_c}\partial_t \delta\psi_R+e^{2g_c}\left(q_e^2 \frac{a_c^2}{U_c}-2\partial_{|\psi|^2 }V_c-\partial_{|\psi|^2 }\tau_c F^2_c\right)\delta\psi_I=0\,.
\end{align}
We will now focus on a time-dependent perturbation of the form,
\begin{align}
    \delta\psi_{R/I}(r,t)=e^{-i\omega(t+S(r))}\delta\tilde{\psi}_{R/I}(r)\,,
\end{align}
which behaves near the boundary as,
\begin{align}
\delta\tilde{\psi}_{R/I}=\frac{\delta\tilde{\psi}^s_{R/I}}{r+R}+\frac{\delta\tilde{\psi}^v_{R/I}}{(r+R)^2}+\cdots\,,
\end{align}
and near the horizon,
\begin{align}
    \delta\tilde{\psi}_{R/I}=\delta\tilde{\psi}_{R/I}^{(0)}+\cdots\,.
\end{align}

The goal for the rest of this section is to extract from \eqref{eq:Critic_eqs} the vevs $\delta\tilde{\psi}^v_{R/I}$ in terms of the sources $\delta\tilde{\psi}^s_{R/I}$, in the small frequency limit. The retarded Green's functions will then follow from the definitions, $G_R(\psi_R,\psi_R)\equiv\frac{\delta\tilde{\psi}^v_{R}}{\delta\tilde{\psi}^s_{R}}$, etc.. It is important to note that we choose our sources to be $\delta\tilde{\psi}_{R/I}^s=\mathcal{O}(\omega^0)$.

A useful observation is that, in the static limit of \eqref{eq:Critic_eqs}, the real and imaginary parts of $\delta\psi$ decouple and satisfy the same equation, which is precisely the equation of motion for the critical mode $\delta\rho_{\ast (0)}$. Making use of this, we can manipulate \eqref{eq:Critic_eqs} to the following form,
\begin{align}\label{eq:Crit_simp}
    \partial_r P_R-i \omega S' P_R+\omega^2 \frac{e^{2g_c}}{U_c}\delta\rho_{\ast (0)}\delta\tilde{\psi}_R-i\omega\,2q_e e^{2g_c}\frac{a_c}{U_c}\delta\rho_{\ast(0)}\delta\tilde{\psi}_I=0\,,\nn
   \partial_r P_I-i \omega S' P_I+\omega^2 \frac{e^{2g_c}}{U_c}\delta\rho_{\ast (0)}\delta\tilde{\psi}_I+i\omega\,2q_e e^{2g_c}\frac{a_c}{U_c}\delta\rho_{\ast(0)}\delta\tilde{\psi}_R=0\,,
\end{align}
with 
\begin{align}
P_{R/I}=e^{2g_c}U_c\left(\delta\tilde{\psi}'_{R/I}\delta\rho_{\ast(0)}-\delta\tilde{\psi}_{R/I} \delta\rho'_{\ast(0)}-i\omega S' \delta\rho_{\ast(0)} \delta\tilde{\psi}_{R/I}\right)\,.
\end{align}

Looking carefully at these equations one immediately deducts that the fields can be expanded in powers of $\omega$ as
\begin{align}
    \delta\tilde{\psi}_{R/I}=\frac{1}{\omega}(\delta\tilde{\psi}_{R/I}^{[0]}+\omega\, \delta\tilde{\psi}_{R/I}^{[1]}+\mathcal{O}(\omega^2))\,,
\end{align}
with the leading part being just a constant multiple of the critical mode, namely 
\begin{align}
\delta\tilde{\psi}_{R/I}^{[0]}=\kappa_{R/I}\delta\rho_{\ast(0)}\,.
\end{align}

We can now solve \eqref{eq:Crit_simp} order by order in $\omega$. At order $\omega^0$, integrating \eqref{eq:Crit_simp} from the horizon to the boundary we get
\begin{align}\label{eq:kappas}
  \delta\rho_  {\ast(0)}^v\delta\tilde{\psi}_R^s+i e^{2g_c^0}\kappa_R\left(\delta\rho_  {\ast(0)}^{(0)}\right)^2-i\kappa_I \frac{\Delta\varrho_{h(2)}}{q_e}=0\,,\nn
   \delta\rho_  {\ast(0)}^v\delta\tilde{\psi}_I^s+i e^{2g_c^0}\kappa_I\left(\delta\rho_  {\ast(0)}^{(0)}\right)^2+i\kappa_R \frac{\Delta\varrho_{h(2)}}{q_e}=0\,.
\end{align}
Integrating the same equations from the horizon to a point $r$ in the bulk instead, one finds
\begin{align}\label{eq:psi_eqs_inter}
\partial_r\left(\frac{\delta\tilde{\psi}_R^{[1]}}{\delta\rho_{\ast(0)}}\right)-iS'\kappa_R+\frac{i}{e^{2g_c}U_c \delta\rho_{\ast(0)}^2 }\left(\kappa_R e^{2g_c^0}\left(\delta\rho_  {\ast(0)}^{(0)}\right)^2-\frac{\kappa_I}{q_e}(\delta \Phi_{b(2)}-\delta\varrho_{\ast h(2)})\right)=0\,,\nn
\partial_r\left(\frac{\delta\tilde{\psi}_I^{[1]}}{\delta\rho_{\ast(0)}}\right)-iS'\kappa_I+\frac{i}{e^{2g_c}U_c \delta\rho_{\ast(0)}^2 }\left(\kappa_I e^{2g_c^0}\left(\delta\rho_  {\ast(0)}^{(0)}\right)^2+\frac{\kappa_R}{q_e}(\delta \Phi_{b(2)}-\delta\varrho_{\ast h(2)})\right)=0\,.
\end{align}
Further integrating \eqref{eq:psi_eqs_inter} from the horizon to the boundary, we arrive at,
\begin{gather}\label{eq:psi_v_R}
    R\, \delta\tilde{\psi}_R^s+\delta\tilde{\psi}_R^{[1]v}-\frac{\delta\rho_{\ast(0)}^v}{\delta\rho_{\ast(0)}^{(0)}}\delta\tilde{\psi}_R^{[1](0)}+\\\notag \int^\infty_0 dr\left(\delta\tilde{\psi}_R^s-iS'\kappa_R \delta\rho_{\ast(0)}^v+\frac{i \delta\rho_{\ast(0)}^v}{e^{2g_c}U_c\delta\rho_{\ast(0)}^2}\left(e^{2g^0_c}\kappa_R \left(\delta\rho_  {\ast(0)}^{(0)}\right)^2-\kappa_I\, \delta \tilde{\Phi}_{b(2)}\right)\right)=0\,,
\end{gather}

\begin{gather}\label{eq:psi_v_I}
    R\, \delta\tilde{\psi}_I^s+\delta\tilde{\psi}_I^{[1]v}-\frac{\delta\rho_{\ast(0)}^v}{\delta\rho_{\ast(0)}^{(0)}}\delta\tilde{\psi}_I^{[1](0)}+\nn \int^\infty_0 dr\left(\delta\tilde{\psi}_I^s-iS'\kappa_I \delta\rho_{\ast(0)}^v+\frac{i \delta\rho_{\ast(0)}^v}{e^{2g_c}U_c\delta\rho_{\ast(0)}^2}\left(e^{2g^0_c}\kappa_I \left(\delta\rho_  {\ast(0)}^{(0)}\right)^2+\kappa_R\,\delta \tilde{\Phi}_{b(2)}\right)\right)=0\,,
\end{gather}
with $\delta\tilde{\Phi}_{b(2)}=\frac{1}{q_e}\left(\delta\Phi_{b(2)}-\delta\varrho_{\ast h(2)}\right).$ Furthermore, at order $\omega$, integrating \eqref{eq:Crit_simp} from the horizon to the boundary and then employing \eqref{eq:psi_eqs_inter} and \eqref{eq:kappas} we find
\begin{gather}\label{eq:hor_psi_1}
i e^{2g^0_c}\delta\tilde{\psi}_R^{[1](0)}\delta\rho_  {\ast(0)}^{(0)}-\frac{i}{q_e}\Delta\varrho_{h(2)}\frac{\delta\tilde{\psi}_I^{[1](0)}}{\delta\rho_  {\ast(0)}^{(0)}}+\nn
\int^\infty_0 dr\left(-i S'  \delta\rho_  {\ast(0)}^v\delta\tilde{\psi}_R^s+\kappa_R\frac{e^{2g_c}}{U_c}\delta\rho_  {\ast(0)}^2+\frac{\delta \Phi_{b(2)}-\delta\varrho_{\ast (2)}}{q_e e^{2g_c} U_c \delta\rho_  {\ast(0)}^2}\left(\kappa_I\, e^{2g^0_c} \left(\delta\rho_{\ast(0)}^{(0)}\right)^2+\kappa_R\,\delta \tilde{\Phi}_{b(2)}\right)\right)=0\,,
\end{gather}

\begin{gather}\label{eq:hor_psi_2}
i e^{2g^0_c}\delta\tilde{\psi}_I^{[1](0)}\delta\rho_  {\ast(0)}^{(0)}+\frac{i}{q_e}\Delta\varrho_{h(2)}\frac{\delta\tilde{\psi}_R^{[1](0)}}{\delta\rho_  {\ast(0)}^{(0)}}+\\\notag
\int^\infty_0 dr\left(-i S'  \delta\rho_  {\ast(0)}^v\delta\tilde{\psi}_I^s+\kappa_I\frac{e^{2g_c}}{U_c}\delta\rho_  {\ast(0)}^2-\frac{\delta \Phi_{b(2)}-\delta\varrho_{\ast (2)}}{q_e e^{2g_c} U_c \delta\rho_  {\ast(0)}^2}\left(\kappa_R\, e^{2g^0_c} \left(\delta\rho_{\ast(0)}^{(0)}\right)^2-\kappa_I\delta \tilde{\Phi}_{b(2)}\right)\right)=0\,.
\end{gather}

Finally, solving \eqref{eq:psi_v_R}-\eqref{eq:hor_psi_2} for the vevs, we arrive at
\begin{align}
    \delta\tilde{\psi}_R^{[1]v}&=\frac{\mathrm{Re}(Z_\pi)}{2}\delta\tilde{\psi}_R^s-\frac{\mathrm{Im}(Z_\pi)}{2}\delta\tilde{\psi}_I^s\,,\nn
    \delta\tilde{\psi}_I^{[1]v}&=\frac{ \mathrm{Im}(Z_\pi)}{2}\delta\tilde{\psi}_R^s+\frac{\mathrm{Re}(Z_\pi)}{2}\delta\tilde{\psi}_I^s\,,
\end{align}
with $Z_\pi$ as given in \eqref{eq:ReZ_pi},\,\eqref{eq:ImZ_pi}. Solving also \eqref{eq:kappas} for the $\kappa$'s, we can read the retarded two-point functions in the small frequency limit\footnote{As a cross-check, we highlight that the Onsager relations are satisfied.},
\begin{align}\label{eq:Greens_Tc}
    G_R(\psi_R,\psi_R)&= G_R(\psi_I,\psi_I)\approx i \frac{\mathrm{Re}(\Gamma_0)}{\omega}+\frac{\mathrm{Re}(Z_\pi)}{2}+\mathcal{O}(\omega)\,,\nn
    G_R(\psi_R,\psi_I)&=- G_R(\psi_I,\psi_R)\approx  i \frac{\mathrm{Im}(\Gamma_0)}{\omega}-\frac{\mathrm{Im}(Z_\pi)}{2}+\mathcal{O}(\omega)\,.
\end{align}

The final results \eqref{eq:Greens_Tc} are, of course, in complete agreement with the two-point functions obtained from \eqref{eq:ord_linear}, if we drop the spatial dependence, set all the sources but $\delta s_\psi$ to zero and take the limit $T\to T_c$.



\section{Conformal invariance}\label{app:Conform}

Starting from \eqref{eq:Transp_final}, our goal here is to show that $Z_2=-1$ for a conformal system, in which the neutral scalar field is trivial, everywhere in the bulk. We consider a family of backgrounds of the form \eqref{eq:background}, labelled by $s,\varrho,\rho_v$. Using a scaling argument for the background solution, similar to the one in section (2.4) of \cite{Kailidis:2023ntq}, we can see that the horizon charge density and the value of the amplitude at the horizon satisfy,
\begin{align}
\varrho_h(s,\varrho,\rho_v)=\varrho\, f_1\left(\frac{s}{\varrho},\frac{\rho_v}{\varrho}\right),\quad \rho^{(0)}(s,\varrho,\rho_v)=f_2\left(\frac{s}{\varrho},\frac{\rho_v}{\varrho}\right)
\end{align}
for some functions $f_1,\,f_2$.
Using these relations, we find,
\begin{align}
s\,\partial_s\varrho_h+\varrho\,\partial_\varrho \varrho_h=\varrho_h-\rho_v\,\partial_{\rho_v}\varrho_h\,,\nn
s\,\partial_s\rho^{(0)}+\varrho\,\partial_\varrho \rho^{(0)}=-\rho_v\,\partial_{\rho_v}\rho^{(0)}\,.
\end{align}
The above formulas, along with \eqref{eq:susc_oe}, then lead to $Z_2=-1+\mathcal{O}(\ves^2)$.

\newpage
\bibliographystyle{utphys}
\bibliography{refsthesis}{}
\end{document}